\documentclass[fleqn,10pt]{article}
\usepackage{fullpage}
\usepackage{amsfonts}
\usepackage{amsmath}
\usepackage{multirow}
\usepackage{longtable}
\usepackage{graphicx}
\usepackage{sidecap}
\usepackage{subcaption}
\usepackage[affil-it]{authblk}
\usepackage{hyperref}
\usepackage{cite}
\usepackage{color}

\newcommand{\eref}[1]{Eq.~(\ref{#1})}
\newcommand{\esref}[2]{Eqs.~(\ref{#1})-(\ref{#2})}

\newcommand{\fref}[1]{Fig.~\ref{#1}}

\def\kv{\mathbf{k}}
\def\mv{\mathbf{m}}
\def\wv{\mathbf{w}}
\def\0v{\mathbf{0}}
\def\qk{q_{\kv}}
\def\qm{q_{\mv}}
\def\ej{\mathbf{e}_j}
\def\ei{\mathbf{e}_i}
\def\Pk{P(k)}
\def\Pw{P(w)}
\def\Pkv{P({\kv})}
\def\Fkm{F_{\kv, \mv}}

\def\fkm{f(\kv, \mv)}

\def\skm{s_{\kv, \mv}}
\def\skmej{s_{\kv, \mv - \ej}}
\def\skmei{s_{\kv, \mv - \ei}}
\def\dskm{\dot{s}_{\kv, \mv}}
\def\ikm{i_{\kv, \mv}}
\def\ikmej{i_{\kv, \mv - \ej}}

\def\nuj{\nu_j}
\def\nui{\nu_i}

\def\nuv{\boldsymbol{\nu}}
\def\dnuj{\dot{\nu}_j}
\def\dnui{\dot{\nu}_i}
\def\drho{\dot{\rho}}
\def\summ{\sum_{\mv}}
\def\sumkk{\sum_{k, \kv}}
\def\sumkm{\sum_{\kv, \mv}}
\def\sumkkm{\sum_{k, \kv, \mv}}
\def\sumj{\sum_{j=1}^n}
\def\sumi{\sum_{i=1}^n}
\def\sumLess{\sum_{\qm < \phi \qk}}
\def\sumMore{\sum_{\qm \geq \phi \qk}}
\def\prodj{\prod_{j=1}^n}
\def\prodi{\prod_{i=1}^n}
\def\prodinj{\prod_{i \neq j}^n}

\def\bj{\beta^s_j}
\def\bi{\beta^s_i}
\def\bij{\beta^i_j}

\def\Bkmj{B_{k_j, m_j}}
\def\Bkmi{B_{k_i, m_i}}

\def\Bkk1{B_{k, k_1}}
\def\Bkomj{B_{k_j - 1, m_j}}
\def\Bkmoj{B_{k_j, m_j - 1}}

\def\gam{\delta} 

\begin{document}

\title{\textbf{Threshold driven contagion on weighted networks}}
\date{\vspace{-5ex}}

\author[1]{Samuel Unicomb}
\affil[1]{\normalsize{Univ Lyon, ENS de Lyon, Inria, CNRS, UCB Lyon 1, LIP UMR 5668, IXXI, F-69342, Lyon, France}}

\author[2,3]{Gerardo I\~niguez}
\affil[2]{\normalsize{Instituto de Investigaciones en Matem\'aticas Aplicadas y en Sistemas, Universidad Nacional Aut\'onoma de M\'exico, 04510 Ciudad de M\'exico, Mexico}}
\affil[3]{\normalsize{Department of Computer Science, School of Science, Aalto University, 00076, Finland}}

\author[1]{M\'arton Karsai \thanks{Corresponding author: marton.karsai@ens-lyon.fr}}

\maketitle

\begin{abstract}
Weighted networks capture the structure of complex systems where interaction strength is meaningful. This information is essential to a large number of processes, such as threshold dynamics, where link weights reflect the amount of influence that neighbours have in determining a node's behaviour. Despite describing numerous cascading phenomena, such as neural firing or social contagion, threshold models have never been explicitly addressed on weighted networks. We fill this gap by studying a dynamical threshold model over synthetic and real weighted networks with numerical and analytical tools. We show that the time of cascade emergence depends non-monotonously on weight heterogeneities, which accelerate or decelerate the dynamics, and lead to non-trivial parameter spaces for various networks and weight distributions. Our methodology applies to arbitrary binary state processes and link properties, and may prove instrumental in understanding the role of edge heterogeneities in various natural and social phenomena.
\end{abstract}

\section*{Introduction}

Weighted networks provide meaningful representations of the architecture of a large number of complex systems where interacting entities, represented as nodes in a graph, are connected with links weighted by the strength of their interactions. Weighted networks are ubiquitous in biological~\cite{horvath2011weighted}, ecological~\cite{luczkovich2003defining}, infrastructure~\cite{barrat2004architecture,Pastor2007Evolution}, social~\cite{Granovetter1973The,wasserman1994social,onnela2007analysis}, information, and economic~\cite{hidalgo2007product} systems, just to mention a few. Their analysis has been in focus from the early stages of complex networks research~\cite{Newman2004Analysis,Wang2005General}, with several measures~\cite{Serrano2009Extracting,Opsahl2009Clustering,Opsahl2010Node} and models~\cite{Barrat2004Weighted,Yook2001Weighted} introduced. These studies show that link weights in real networks are usually heterogeneous, may be correlated with the network structure~\cite{Granovetter1973The,Onnela2007Structure}, and can even capture signed relationships~\cite{szell2010multirelational}. More importantly, weights help to differentiate links of varying importance, influence, and role. On a microscopic level, weights identify the most relevant neighbours of a node~\cite{Saramaki2014Persistence}; on a network level, they indicate links with special roles or positions in the system~\cite{Granovetter1973The,Onnela2007Structure}. Such information is crucial for dynamical processes evolving on weighted networks. Examples can be found in epidemiology, where important ties maintained by frequent interactions may enhance the spread of infection locally, while ties with infrequent interactions but located between densely connected parts of the network may suppress diffusion globally~\cite{Onnela2007Structure,Zhu2017Social}. Link weights are also relevant in phenomena like random walks, spin models, synchronisation, evolutionary games, or even cascading failures. Despite this, weighted networks have been less studied than their unweighted counterparts, especially for threshold driven processes, which are essential in systems of self-organised criticality~\cite{jensen1998self,corral1995self,Bak1993Punctuated}, epidemiology~\cite{joh2009dynamics}, firing neurons~\cite{koch1998methods,stein1967some,gerstner2014neuronal}, or social contagion~\cite{Granovetter1973The,watts2002simple}.

In threshold driven processes, the state of an entity changes when the concentration of incoming stimuli or cumulating force reaches a certain threshold. Typical examples are neural systems~\cite{koch1998methods,stein1967some}, earthquakes~\cite{Herz1995Earthquake}, and solar flares~\cite{boffetta1999power,charbonneau2001avalanche}, commonly identified as self-organised critical systems driven by integrate-and-fire mechanisms. Thresholds play a role in some epidemic diseases, such as tuberculosis and dysentery~\cite{joh2009dynamics}, where infection requires the concentration of pathogens in an individual to overcome a threshold. Moreover, thresholds are also associated with social contagion phenomena, where social influence from acquaintances may change the behaviour of an individual after reaching a cognitive limit. Studies of so-called \emph{complex contagion} date back to Schelling, Axelrod, and Granovetter, but have gained recent interest thanks to a seminal cascade model due to Watts~\cite{watts2002simple}, and thanks to the enormous amount of digital data on human behaviour collected to observe, analyse and model social contagion. In threshold models on networks links are usually considered unweighted, such that the stimuli or influence arriving from each neighbour contributes equally to reaching the behavioural threshold. Although this assumption simplifies their modelling, it does not lead to an accurate representation of real threshold dynamics. For example, in neural systems synaptic connections have weights that quantify the strength of incoming stimuli, and contribute unequally in bringing neurons to an excited state, as recognised recently in models of neural population dynamics~\cite{iyer2013influence}. In social systems link weights are associated with tie strengths that quantify the social influence that individuals have on their peers. Measurement of tie strength is a long standing challenge, but it is generally accepted that social ties are not equal, as some of them are more influential than others on one's decision making. Surprisingly, apart from some recent studies~\cite{kempe2003maximizing,hurd2013watts,cox2016spread}, weights have been commonly overlooked in models of threshold driven phenomena.

Our aim is to close this gap by exploring the effect of weight heterogeneities on threshold driven contagion processes. We first study a dynamical variant of the Watts cascade model on a simple system, a random regular network with a bimodal weight distribution. We then provide an analytical solution of the dynamics, for arbitrary degrees and weights, together with numerical simulations and combinatorial arguments to show that the speed of spreading depends non-monotonously on the strength of weight heterogeneities and may radically accelerate or decelerate as compared to the unweighted case, even for fixed thresholds. We also find this effect under more realistic synthetic scenarios, such as scale-free networks and lognormal weight distributions, as well as in data-driven simulations over large-scale empirical weighted networks. Our contribution is a first step into the largely unexplored modelling of dynamical processes with heterogeneous interactions, typical in neural systems and social contagion. Moreover, our results may have broader implications as our methodology is not specific to threshold dynamics and may be easily extended to any binary state process, while our study and conclusions may be useful in accurately modelling other dynamical phenomena over weighted networks.

\section*{Results}

\begin{figure*}[t]
  \centering
\includegraphics[width=1.0\textwidth]{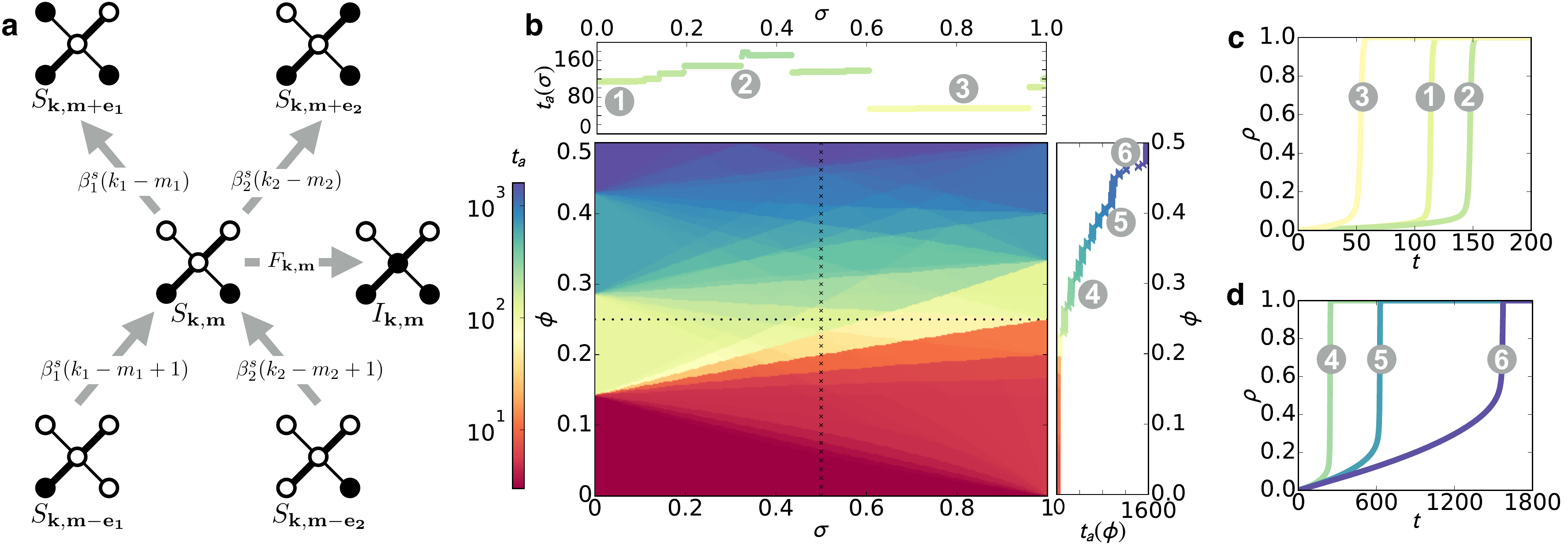}
  \caption{\small\textbf{Threshold driven contagion and cascade evolution on weighted networks.} \textbf{(a)} Transitions into and out of class $s_{\kv,\mv}$ of susceptible nodes in a network with two weights ($n=2$). Susceptible nodes may enter or leave $s_{\kv,\mv}$ with rate $\beta_1^s$, $\beta_2^s$ via the infection of neighbours with weight type $j=1, 2$, or via their own infection with rate $\Fkm$. \textbf{(b)} Parameter dependence of the time $t_a$ of cascade emergence (main panel) on a random regular network with degree $k=7$, and bimodal weight distribution with mean $\mu=1$ and standard deviation $\sigma$ (for further details see text). Cascade speed is measured by the time $t_a$ to reach $75\%$ infection. For fixed threshold $\phi$ and varying $\sigma$, $t_a$ changes non-monotonously, while for fixed $\sigma$ and varying $\phi$, dynamics slows down for increasing $\phi$ (top/right panels, corresponding to horizontal/vertical dashed lines in main panel). \textbf{(c-d)} Spreading time series $\rho (t)$ for selected parameter values in (b). Simulation results in  (b-d) are averages of $25$ simulations with $p=2\times 10^{-4}$ and $N=10^4$.}
\label{fig:1}
\end{figure*}

\subsection*{Threshold model and approximate solutions}

To study threshold driven dynamical processes over weighted networks we build on a seminal model proposed by Watts~\cite{watts2002simple}. Following its standard formulation~\cite{watts2002simple,singh2013threshold,ruan2015kinetics,karsai2016local}, we define a monotone binary-state dynamics over a weighted, undirected network of size $N$. Degrees take discrete values $k = 0, \ldots, N-1$ according to the distribution $\Pk$, and edge weights $w > 0$ are continuous variables with distribution $\Pw$. The edge weight $w_{ij}$ represents the capacity of connected nodes $i$ and $j$ to influence each other. Accordingly, the node strength $q_k(i) = \sum_{j=1}^k w_{ij}$ is the total influence node $i$ receives from its $k$ neighbours. Like in other conventional models of spreading dynamics~\cite{porter2016dynamical}, nodes can be in two mutually exclusive states, susceptible (initially all nodes), or infected (also called adopter in the social contagion literature). A susceptible node can become infected either spontaneously with rate $p$~\cite{ruan2015kinetics,karsai2016local}, or if the influence of its infected neighbours exceeds a given threshold $\phi$ ($0 < \phi < 1$). However, influence may vary from neighbour to neighbour. We implement this idea by defining the \emph{partial strength} $q_m(i) = \sum_{j=1}^m w_{ij}$ associated with the influence of the $m$ infected neighbours on node $i$ (where $0 \leq m \leq k$). If the condition $q_m \geq \phi q_k$ is fulfilled, node $i$ becomes infected and remains so indefinitely. For simplicity we assume that all nodes have the same threshold $\phi$, just like in many other studies~\cite{watts2002simple,singh2013threshold} (implementation details in Methods).

We explore this model analytically by extending Gleeson's approximate master equation (AME) formalism for stochastic binary-state dynamics~\cite{porter2016dynamical,gleeson2013binary,gleeson2008cascades,gleeson2011high} over weighted networks. Although we only consider monotone dynamics in detail, our formalism can easily be extended to arbitrary binary state processes (see Supplementary Information [SI]). The original AME formalism considers unweighted networks with an arbitrary degree distribution but are otherwise maximally random. It assumes that all nodes with degree $k$ and number of infected neighbours $m$ follow the same dynamics, forming a node class $(k, m)$ that can be described by a single pair of rate equations. In order to extend this formalism to weighted networks, we discretise $\Pw$ and assume only $n$ possible {\it weight types} $w_j$, such that all distinct weights in the network are contained in the {\it weight vector} $\wv = (w_1, \ldots, w_n)^{\mathrm{T}}$. Then, a node in class $(k, m)$ has $k_j$ links with weight $w_j$ and $m_j = 0, \ldots, k_j$ infected neighbours across these links, such that $k = \sumj k_j$ and $m = \sumj m_j$. We can further define a {\it degree vector} $\kv = (k_1, \ldots, k_n)^{\mathrm{T}}$ and a {\it partial degree vector} $\mv = (m_1, \ldots, m_n)^{\mathrm{T}}$, generalising the strength and partial strength to $\qk = \kv \cdot \wv$ and $\qm = \mv \cdot \wv$. Nodes in class $(\kv, \mv)$ have identical strengths and partial strengths, and thus follow the same pair of rate equations for the fraction $\skm (t)$ [$\ikm (t)$] of $\kv$-nodes that are susceptible (infected) at time $t$ and have partial degree vector $\mv$ (see Methods and SI).
 
In threshold driven contagion a susceptible node can become infected in two ways, either spontaneously with rate $p$, or if its weighted threshold $\phi$ is reached. Then, the infection rate of susceptible nodes in class $(\kv, \mv)$ is
\begin{equation}
\label{eq:thresRule}
\Fkm =
\begin{cases}
p & \quad \qm < \phi \qk \\
1 & \quad \qm \geq \phi \qk
\end{cases}, \quad k > 0,
\end{equation}
with $F_{\0v, \0v} = p$. The stepwise nature of $\Fkm$ allows us to map the rate equations for $\skm$ and $\ikm$ to a reduced-dimension system, as has been done previously for the Watts threshold model~\cite{porter2016dynamical,gleeson2013binary,gleeson2011high} and unweighted complex contagion~\cite{ruan2015kinetics,karsai2016local}. Namely, if we consider as aggregated variables the density $\rho(t)$ of infected nodes and the probability $\nuj(t)$ that a randomly chosen neighbour (across a $j$-type edge) of a susceptible node is infected (for definitions see Methods), then the description of the dynamics can be reduced to the system of $n + 1$ equations
\begin{subequations}
\label{eq:reducedAMEs}
\begin{align}
\dnuj &= g_j(\nuv, t) - \nuj, \\
\drho &= h(\nuv, t) - \rho,
\end{align}
\end{subequations}
where $\nuv = (\nu_1, \ldots, \nu_n)^{\mathrm{T}}$ is the vector of probabilities $\nuj$ for all weight types, and $g_j(\nuv, t)$ and $h(\nuv, t)$ are functions of binomial terms (see Methods and SI).

\subsection*{Regular networks with bimodal weights}

To study the dynamics of our model we first consider a simple structure, the configuration-model $k$-regular network ($k=7$). Edge weights are sampled from a bimodal distribution with $n=2$ values, denoted strong ($w_1$) and weak ($w_2$). The weight distribution is characterised by its average $\mu$, standard deviation $\sigma \geq 0$, and the fraction $\delta$ of strong links. Thus weights take the values $w_1 = \mu + \sigma \sqrt{ (1 - \delta) / \delta }$ and $w_2 = \mu - \sigma \sqrt{ \delta / (1 - \delta) }$. The parameter $\delta$ contributes to the skewness of $\Pw$, initially fixed to the symmetric case $\delta = 0.5$. The parameter $\sigma$ interpolates weight heterogeneity between the homogeneous case of an unweighted network ($\sigma=0$), and the most heterogeneous case of a diluted network ($\sigma = \mu \sqrt{ (1 - \delta) / \delta }$), where only strong links have influence and the weak are functionally absent. After fixing the spontaneous infection rate $p$ and skewness $\delta$, our model has only two parameters, $\sigma$ and $\phi$ (Fig.~\ref{fig:1}b). Similar to other dynamical cascade models \cite{ruan2015kinetics,karsai2016local},  contagion initially evolves at a linear rate close to $p$ until the density $\rho(t)$ of infected nodes reaches a critical value, triggering a rapid cascade of infection that spreads through the whole network (sample scenarios in Fig.~\ref{fig:1}c-d). Thus, to characterise the speed of dynamics we introduce the quantity $t_a$, the time when infection density reaches a set value ($\rho=0.75$), called the absolute time of cascade emergence. We measure $t_a$ via numerical simulations of the $(\sigma, \phi)$-parameter space (Fig.~\ref{fig:1}b), which shows unexpected dependencies on both parameters. On one hand, for fixed $\sigma$ and increasing $\phi$ the dynamics slows down, since nodes with higher thresholds require more infected neighbours to become infected. On the other, for fixed $\phi$ the dynamics depends \emph{non-monotonously} on $\sigma$, where cascades may evolve either faster or slower as we increase weight heterogeneity, relative to the unweighted case ($\sigma=0$).

\begin{figure*}[t]
  \centering
  \includegraphics[width=1.0\textwidth]{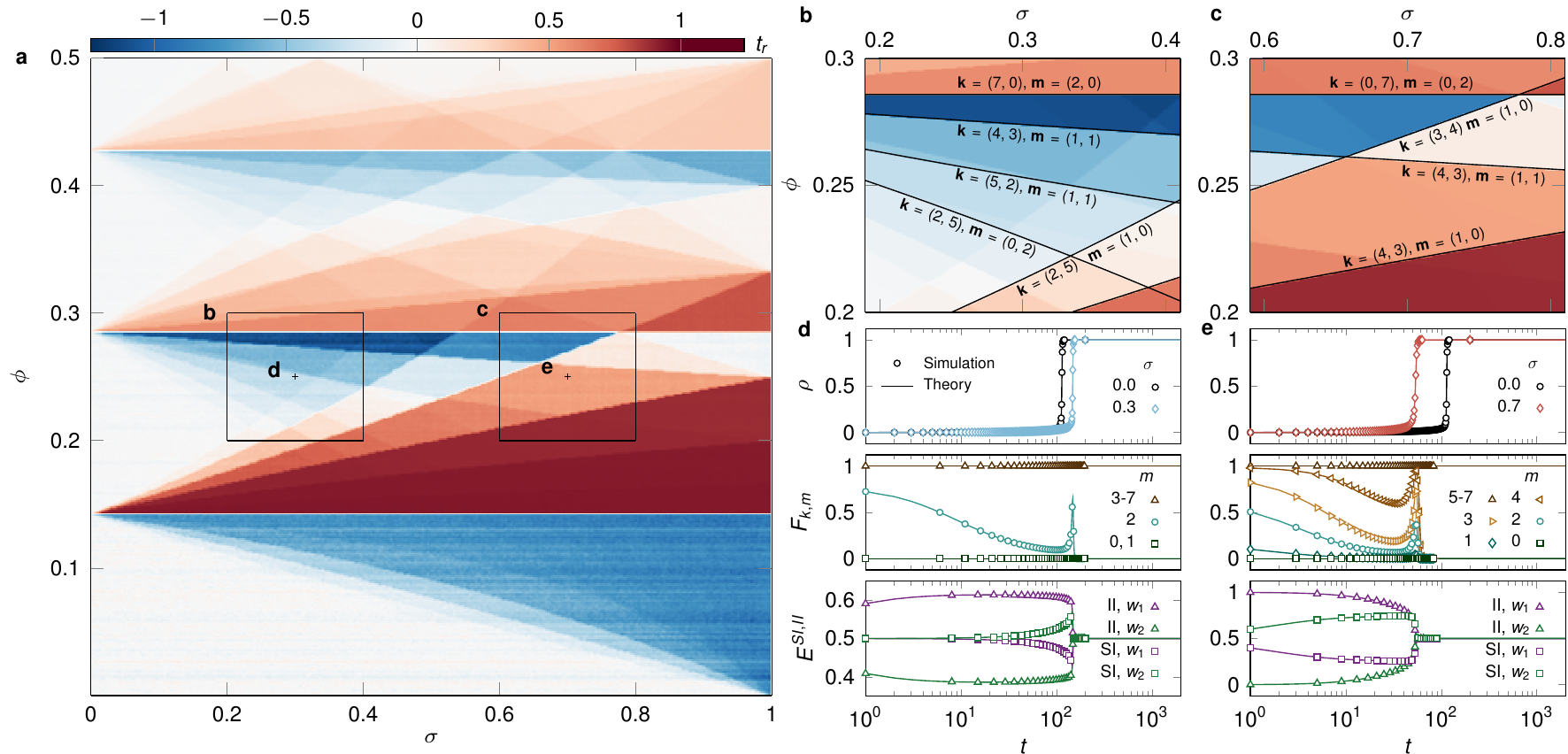}
  \caption{\small \textbf{Relative speed of threshold driven cascades on weighted networks.} \textbf{(a)} Relative time $t_r$ of cascade emergence on $(\sigma, \phi)$-parameter space, simulated over $k$-regular regular networks ($k=7$) with $\mu = 1$, $\delta = 0.5$, $p = 2\times 10^{-4}$, $N = 10^4$ and averaged over 25 realisations. Time of cascades for given $\phi$ is either higher or lower than the corresponding case $(0, \phi)$ of an unweighted network. \textbf{(b-c)} Selected regions of parameter space in (a), where $t_r$ is instead calculated from the numerical solution of the AME systems in Eq.~\ref{eq:reducedAMEs}. Boundaries are obtained from a combinatorial argument (see Methods and SI) for various $( \kv, \mv )$ classes. For example, the boundary $\kv = (2,5)$, $\mv = (1,0)$ separates networks where nodes with $k_1 = 2$ strong links and $k_2 = 5$ weak links may (or may not) be infected by $m_1 = 1$ strong infected neighbour. \textbf{(d-e)} Quantities characterising the dynamics in simulations (symbols) and AMEs (lines) for $\phi=0.25$ and $\sigma$ corresponding to the unweighted case, as well as to a slow (d) or fast (e) cascade. Quantities are the infection density $\rho(t)$ (upper panel), aggregated infection rate $F_{k, m} (t)$ for various numbers of infected neighbours $m$ (middle panel), and fractions of strong ($w_1$) and weak ($w_2$) links inside the infected cluster [$E^{II} (t)$] and on the surface of it [$E^{SI} (t)$] (bottom panel). Simulation and theory results in (a-e) agree perfectly.}
    \label{fig:2}
\end{figure*}

We concentrate on the $\sigma$ dependency by calculating $t_r =[ t_a(0,\phi) - t_a(\sigma,\phi) ] / t_a(0,\phi)$, the time of cascade emergence relative to the unweighted case with the same $\phi$ value. (Fig.~\ref{fig:2}a). The $(\sigma, \phi)$-parameter space for $t_r$ is highly structured and driven by competing effects of key $(\textbf{k},\textbf{m})$ classes, which either reduce or enhance the speed of the spreading process as compared to the unweighted case. We also explore the corresponding numerical solution of the AME systems in Eq.~\ref{eq:reducedAMEs}, as well as an independent combinatorial solution for the boundaries between regions of low and high cascade speed (Fig.~\ref{fig:2}b-c) (see Methods and SI). Both the AME and combinatorial solutions perfectly recover the parameter space obtained by simulations. To further explore how weight heterogeneities produce slow or fast cascades, we partition the system according to the number $m$ of infected neighbours required for infection, and measure the aggregated infection rate $F_{k,m}(t) = \sumkm \Pkv \Fkm \skm(t) / \sumkm \Pkv \skm(t)$ and other determinant quantities in several spreading scenarios (Fig.~\ref{fig:2}d-e).

In the neutral scenario, all $(\kv, \mv)$ classes of the weighted network share the same dynamics as the corresponding $(k,m)$ class in an unweighted network, so $F_{k,m} = p$ or $1$ and weights have no impact on contagion, meaning $t_r = 0$. In a decelerating scenario like $\phi=0.25$ and $\sigma=0.3$ (Fig.~\ref{fig:2}d), $F_{k, m}$ for any $m$ is equal to its unweighted counterpart, except for the $m = 2$ class, whose adoption rate is $1$ in the unweighted case but it is strongly suppressed in the weighted case, and thus decreases the overall spreading speed. At the same time for a high-speed scenario, like $\phi=0.25$ and $\sigma=0.7$, competing effects from several $(\kv, \mv)$ classes determine the dynamics (Fig.~\ref{fig:2}e). The rate $F_{k, m}$ for $m = 2, \dots, 4$ is lower than 1 which is a decelerating effect (as in the previous case), but the rate $F_{k, 1}$, which is equal to $p$ in the unweighted case, is larger than $p$, and since at the early stages of contagion the number of nodes in class $m=1$ is larger than in any other class with $m > 1$, spreading evolves rapidly to an early cascade. 

Furthermore, an asymmetry is observed to emerge in the fractions of weak and strong links connecting infected [$E^{II} (t)$] or susceptible and infected [$E^{SI} (t)$] nodes (see Methods). Since strong ties contribute the most in reaching the threshold of a node, they participate earlier in the contagion and comprise most ties in the infected subgraph. Conversely, weak ties dominate the surface of the cascade by connecting infected and susceptible nodes. This asymmetry in edge types is an essential feature of weighted contagion that is trivially absent in the unweighted case. This asymmetry evolves both in cases of accelerated and decelerated spreading, with amplitude dependent on the absolute value of the relative speed of contagion. Note that results from simulations (symbols in Fig.~\ref{fig:2}d and e) and AMEs (lines in Fig.~\ref{fig:2}d and e) agree very well, for all quantities studied.

\begin{figure}[t]
  \centering
  \includegraphics[width=0.5\textwidth]{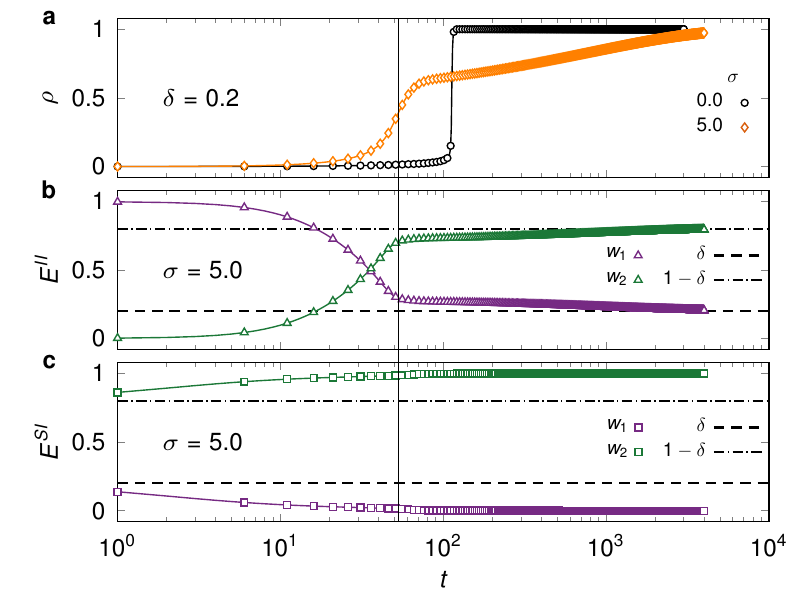}
  \caption{\small \textbf{Effect of skewed weight distributions on cascade evolution.} \textbf{(a)} Infection density $\rho (t)$ on $k$-regular networks ($k=7$) and a bimodal weight distribution with $\mu = 3$ and $\delta=0.2$, both for unweighted ($\sigma = 0$) and heterogeneous ($\sigma > 0$) cases. \textbf{(b-c)} Fractions of strong ($w_1$) and weak ($w_2$) links connecting two infected nodes in the bulk of the infected component [$E^{II} (t)$, b] and susceptible and infected nodes on its surface [$E^{SI} (t)$, c] in the heterogeneous spreading scenario of (a). Simulations (symbols) are averaged over 25 realisations with $p=2 \times 10^{-4}$ and $N=10^4$, and compared with the corresponding AME solution of \esref{eq:sirates}{eq:reducedAMEs} (lines). Dashed lines are the expected fractions of weak and strong links as determined by $\delta$, and the vertical line shows the inflection point of $\rho$ in the heterogeneous case of (a), which coincides with a turning point of $E^{II}$ in (b).}
  \label{fig:3}
\end{figure}

Up until now we have considered the symmetric case $\delta = 0.5$ with equal numbers of strong and weak links. However, by skewing the weight distribution we observe an additional effect of weight heterogeneities on the spreading behaviour. When $\delta = 0.2$ the extent of the cascade decreases for large $\sigma$ with respect to the unweighted case (Fig.~\ref{fig:3}a). In this case, despite their sparsity, strong links again drive the contagion, but are soon exhausted causing spreading to slow down and continue via spontaneous or infrequent threshold driven infections over weak ties (Fig.~\ref{fig:3}b). Indeed, strong links dominate the bulk of the infected component, but disappear quickly from its surface (Fig.~\ref{fig:3}c). These so-called {\it partial cascades}, which do not infect the whole system through the cascade, are associated with skewness and a sufficiently large standard deviation in the weight distribution and are reminiscent of the slow spreading caused by immune nodes, as well as low connectivity networks in unweighted complex contagion~\cite{ruan2015kinetics,karsai2016local,watts2002simple}. Overall, we identify non-monotonous spreading behaviour and partial cascades as the main consequences of weight heterogeneities in threshold driven contagion.

\subsection*{Heterogeneous synthetic and real networks}

\begin{figure*}[t]
  \centering
\includegraphics[width=1.0\textwidth]{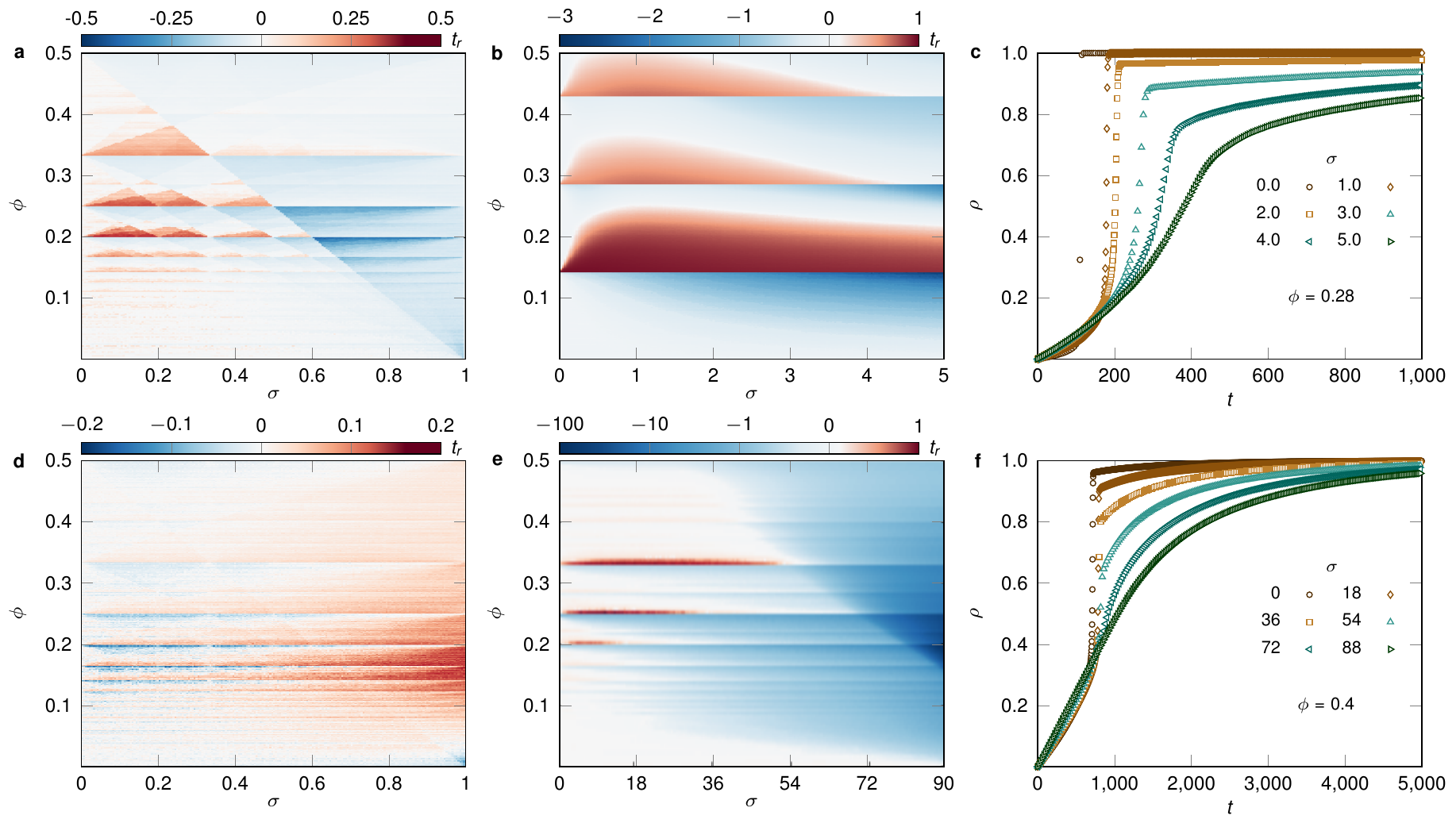}
  \caption{\small \textbf{Threshold contagion on heterogeneous synthetic and real networks.} \textbf{(a)} Relative time $t_r$ of cascade emergence on $(\sigma, \phi)$-parameter space, simulated over synthetic scale-free networks with degree exponent $\tau = 2.5$, average degree $z = 4.54$ and minimum degree $k_{\mathrm{min}} = 2$. Link weights are bimodally distributed with $\mu = 1$ and $\delta = 0.5$. \textbf{(b)} Same as (a) but over a $k$-regular network ($k = 7$) and a lognormal weight distribution with $\mu = 1$. \textbf{(c)} Infection density $\rho (t)$ in the lognormal case of (b) for $\phi=0.28$ and varying $\sigma$. The skewness of the weight distribution induces partial cascades in contagion. \textbf{(d)} Relative time $t_r$ of cascade emergence on $(\sigma, \phi)$-parameter space, simulated over a signed social network of Wikipedia editors with heterogeneous degrees and skewed bimodal weight distribution (see Methods). \textbf{(e)} Same as (d) but over a mobile phone call (MPC) network with heterogeneously distributed degrees and weights, and $\mu = 37.7$. \textbf{(f)} Infection density $\rho (t)$ in the MPC network of (e) for $\phi = 0.4$ and varying $\sigma$. Synthetic networks in (a-b) have $N=10^4$ and parameter spaces are averaged over $25$ realisations. Parameter space in (d) is averaged over $10^3$ realisations, while (e) is the result of a single realisation. All simulations correspond to $p = 2\times 10^{-4}$.}
    \label{fig:4}
\end{figure*}

Although regular networks and bimodal weights are useful in explaining the impact of weights in a simple setting, they are rather unrealistic given that real complex networks commonly appear with broad degree and weight distributions~\cite{barrat2004architecture}. Thus, in the following section we address how threshold driven contagion is influenced by weights using simulations in heterogeneous synthetic and real weighted networks (Fig.~\ref{fig:4}). We expect degree heterogeneities to affect threshold driven processes since thresholds are defined relative to the degree (or strength) of nodes. As a first step we take configuration-model generated scale-free networks with degree distribution $P(k)\sim k^{-\tau}$ with exponent $\tau=2.5$, but keep a bimodal weight distribution with $\mu=1$ and $\delta=0.5$ (Fig.~\ref{fig:4}a). The increased number of $( \kv, \mv )$ classes fragment the $(\sigma, \phi)$-parameter space for $t_r$, but its structure still reveals areas of slow and fast cascades and can be understood using the same mechanisms as the $k$-regular case. Real world examples of this synthetic structure are signed social networks, like the network of Wikipedia editors~\cite{WikiSigned}, where edge signs indicate the parity of a social interaction like trust, intimacy, or influence. We simulate our threshold model over this real social network by associating $+$ and $-$ tie signs with strong ($w_1$) and weak ($w_2$) links, thus obtaining a weighted network with $\delta = 0.88$ and arbitrary $\sigma$ (Fig.\ref{fig:4}d) (see Methods). Despite structural correlations, the Wikipedia $(\sigma, \phi)$-parameter space is qualitatively similar to the synthetic scale-free case, although structural correlations and the high $\delta$ modify the areas of relative acceleration and deceleration. To further validate these observations we have also analysed configuration-model random networks and another empirical signed network, the Pardus dataset~\cite{szell2010multirelational} (see SI).
 
Weights in empirical networks are broadly distributed and approximated by scale-free or lognormal distributions, which we address by exploring the threshold model on $k$-regular networks ($k = 7$) and a lognormal weight distribution with average $\mu = 1$ (Fig.~\ref{fig:4}b). Even though all nodes have the same degree, diversity of weight values increases the number of $( \kv, \mv )$ classes, smoothing out the $(\sigma, \phi)$-parameter space with respect to the bimodal case but qualitatively maintaining its non-monotonous patterns of slow and fast cascades. The standard deviation $\sigma$ controls the skewness of the weight distribution and determines the temporal evolution of contagion, promoting partial cascades for large $\sigma$ (Fig.\ref{fig:4}c). Finally, we consider threshold driven contagion in a large empirical weighted social network, an aggregated mobile phone call (MPC) network, where weights are proportional to the number of calls between individuals (Fig.\ref{fig:4}e) (see Methods). This network has broad degree and weight distributions~\cite{onnela2007analysis}, communities, degree correlations and Granovetter-type degree-weight correlations~\cite{Granovetter1973Strength}. Yet, the $(\sigma, \phi)$-parameter space of the MPC network is similar to previous examples, apart from the magnitude of the slowing down effect when weights are strongly heterogeneous. As before, skewness in the weight distribution temporally inhibits contagion and induces partial cascades (Fig.\ref{fig:4}f). Our data-driven simulations show that, even in empirical networks of vastly different origins, threshold driven contagion strongly depends on link weights via simple mechanisms that can be understood by master equations or combinatorial arguments. This dependence may be responsible for the diverse dynamical scenarios of threshold driven contagion observed in nature, like the diffusion of information in techno-social networks, which typically reaches a limited population, but can occasionally unfold globally through slow or fast cascades of adoption.

\section*{Discussion}

In complex networks, weights quantify the strength of interactions between nodes and distinguish neighbours by the relevance or influence among them. Threshold driven contagion in empirical settings is particularly sensitive to link weights, since influence between connected nodes may vary enormously, thus changing the temporal pattern of global spreading. Nevertheless, contagion is commonly studied over unweighted networks where links are considered equal. Our aim in this paper has been to address this shortfall by systematically studying a threshold model on synthetic and empirical weighted networks. We explore networks with increasing complexity, from configuration-model networks with bimodal or lognormal weights, to real world networks with broad degree and weight distributions as well as multiple correlations. We show that threshold driven contagion depends non-monotonously on weight heterogeneity, creating slow or fast cascades relative to the equivalent unweighted spreading process. Via numerical simulations, master equations and combinatorial arguments, we find that this effect is the result of competing configurations of degree, weight, and infected neighbours that slow down or speed up contagion. We also observe that an imbalance in the amount of large and small weights leads to partial cascades, and smoother temporal patterns of spreading than those in unweighted networks. By analysing many degree and weight configurations, we show that these features are systematic and thus may drive a variety of real world contagion phenomena.

Our contribution opens up directions of research in the largely unexplored area of dynamical processes with heterogeneous interactions. First, the weight-based, master equation formalism described here can be modified to consider any interaction quality like direction and type, thus providing analytical tools to characterise threshold driven contagion in temporal and multiplex networks. Second, our methodology may be used to describe any binary-state dynamics and thus a broad class of empirical processes over weighted networks. We expect our results to find meaningful applications in fields where threshold driven contagion is relevant, like computational epidemiology, neural networks, and social contagion. In these fields our modelling framework, which distinguishes the varied roles and influence of links, may lead to breakthroughs in the understanding and prediction of specific temporal features of global pandemics, collective neural firing, or the adoption of innovations and behavioural norms.

\section*{Methods}

\subsection*{Numerical implementation}

We implement weighted complex contagion numerically via Monte Carlo simulations of a monotone binary-state dynamics. Node states change from susceptible to infected in asynchronous random order in a series of time steps. Once a node state changes from susceptible to infected, it remains so for the rest of the dynamics, thus ensuring a frozen final state for the finite system where no more state changes take place. Each time step consists of $N$ node updates. In each node update, a randomly selected node becomes spontaneously infected with probability $p$, or else it adopts only if the weighted threshold rule $q_m \geq \phi q_k$ is satisfied [see Eq.~\ref{eq:thresRule}]. This is the case if the selected node is susceptible; if it is infected, no action is taken. We assume that nodes with $k=0$ receive no influence from the rest of the network (for any value of $\phi$), and therefore can only change state spontaneously. Regarding synthetic networks, we only consider configuration-model networks~\cite{newman2010networks} with an uncorrelated distribution of edge weights on top of them, i.e. an ensemble of networks specified by the distributions $\Pk$ and $\Pw$, but otherwise maximally random. Thus, the distributions $\Pk$ and $\Pw$ (together with $p$) determine the average topological state and dynamical evolution of the system.

\subsection*{AMEs in weighted networks}

The dynamics of our threshold model evolves in small time intervals $dt$. Accordingly, the rate equations for the fractions $\skm (t)$ [$\ikm (t)$] of $\kv$-nodes that are susceptible (infected) at time $t$ and have partial degree vector $\mv$ are
\begin{subequations}
\begin{eqnarray}
\frac{d}{dt} \skm &=& -\Fkm \skm - \sumj \bj (k_j - m_j) \skm + \sumj \bj (k_j - m_j + 1) \skmej \label{eq:skm}, \\
\frac{d}{dt} \ikm &=& +\Fkm \skm - \sumj \bij (k_j - m_j) \ikm + \sumj \bij (k_j - m_j + 1) \ikmej \label{eq:ikm},
\end{eqnarray}
\label{eq:sirates}
\end{subequations}
where $\Fkm$ is the rate of infection of susceptible nodes in class $(\kv, \mv)$, and the other terms quantify the rates at which susceptible nodes leave and enter the class $(\kv, \mv)$ via the infection of susceptible neighbours. The $j$-th basis vector of dimension $n$ is denoted by $\ej$ ($s_{\kv, -\ej} \equiv 0$), while $\bj(t)$ [$\bij(t)$] is the rate at which a $j$-type susceptible neighbour of a susceptible (infected) node becomes infected (see \fref{fig:1}a). The AME system~(\ref{eq:sirates}) applies to all monotone binary-state dynamics over edge-heterogeneous networks, regardless of the form of $\Fkm$, and its solution provides a very accurate description of the dynamics, even if the number of equations to solve grows rapidly with $n$. Moreover, variables in \eref{eq:sirates} satisfy the normalisation condition
\begin{equation}
\label{eq:normCond}
\summ \ikm + \summ \skm = 1.
\end{equation}
If $\kv$ is distributed according to $\Pkv$, the probability that a randomly selected node has degree $k$ and degree vector $\kv$ is $\Pk \Pkv$. Then, the rates $\bj(t)$ and $\bij(t)$ are
\begin{subequations}
\label{eq:rateBs}
\begin{align}
\bj(t) &= \frac{\sumkkm \Pk \Pkv (k_j - m_j) \Fkm \skm(t)}{\sumkkm \Pk \Pkv (k_j - m_j) \skm(t)}, \\
\bij(t) &= \frac{\sumkkm \Pk \Pkv (k_j - m_j) \Fkm \ikm(t)}{\sumkkm \Pk \Pkv (k_j - m_j) \ikm(t)},
\end{align}
\end{subequations}
where the sum over all degrees, strength and partial strength vectors is written explicitly as
\begin{equation}
\label{eq:sumsDef}
\sumkkm \bullet = \sum_{k=k_{min}}^{k_{max}} \sum_{\kv} \sum_{m_1 = 0}^{k_1} \ldots \sum_{m_n = 0}^{k_n} \bullet.
\end{equation}
The second sum runs over all strength vectors $\kv = (k_1, \ldots, k_n)^{\mathrm{T}}$ satisfying the constraint $k = \sumj k_j$.

\paragraph{Aggregated variables and the reduced AMEs.} Variables in \eref{eq:reducedAMEs} are the fraction of infected nodes in the system,
\begin{equation}
\label{eq:fracAdopt}
\rho(t) = 1 - \sumkkm \Pk \Pkv \skm(t),
\end{equation}
and the probability that a randomly chosen neighbour (across a $j$-type edge) of a susceptible node is infected,
\begin{equation}
\label{eq:nuDef}
\nuj(t) = \sumkk \Pk \Pkv \frac{ \summ m_j \skm(t) }{ \summ k_j \skm(t) }.
\end{equation}
Also, \eref{eq:reducedAMEs} includes the functions of binomial terms 
\begin{subequations}
\label{eq:hgFactors}
\begin{align}
g_j(\nuv, t) &= f_t + (1 - f_t) \sumkk \frac{k_j}{z_j} \Pk \Pkv \sumMore \Bkomj(\nuj) \prodinj \Bkmi (\nui), \\
h(\nuv, t) &= f_t + (1 - f_t) \sumkk \Pk \Pkv \sumMore \prodj \Bkmj(\nuj),
\end{align}
\end{subequations}
with $f_t = 1 - (1 - p) e^{-p t}$, $z_j$ the average number of $j$-type edges of a node, and $\Bkmj = \binom{k_j}{m_j} \rho^{m_j} (1 - \rho)^{k_j - m_j}$ the binomial distribution.

\paragraph{Initial conditions.} We assume that at time $t = 0$ there is an infinitesimally small fraction of infected nodes randomly distributed in the network, so the initial condition for \eref{eq:sirates} is
\begin{equation}
\label{eq:iniCondAMEs}
\skm (0) = \prodj \Bkmj (0).
\end{equation}
In the reduced AMEs, \eref{eq:iniCondAMEs} corresponds to $[\nuv(0), \rho(0)] = (\0v, 0)$.

\subsection*{Combinatorial solution of phase boundaries}

Taking the equality in the threshold rule, Eq.~\ref{eq:thresRule}, and writing $q_{\textbf{k}}$, $q_{\textbf{m}}$ explicitly, we obtain $\phi = \mv \cdot \wv / \kv \cdot \wv$, where $\wv$ implicitly depends on $\sigma$. After solving this equation for a given $\kv$ and $\mv$, we associate the solution with a boundary line of $t_{r}$ values in $(\sigma, \phi)$-parameter space (Fig.~\ref{fig:2}b-c). These boundaries separate network configurations where the corresponding $( \kv, \mv )$ class does or does not satisfy the threshold rule. For example, the boundary $\kv = (3, 4)$, $\mv = (1, 0)$ (Fig.~\ref{fig:2}c) separates networks where nodes with $k_1 = 3$ strong links and $k_2 = 4$ weak links may be infected by only $m_1 = 1$ strong infected neighbour. If two networks differ only in the rate of infection of nodes in this $( \kv, \mv )$ class (so that one is eligible for infection and not the other), we observe a difference in spreading time (for details see SI).

\subsection*{Bulk and interface of the contagion cluster}

We characterise the effect of weights in threshold driven contagion by measuring how many $j$-type links, $1 \leq j \leq n$, are on the bulk and at the surface of cascades. Explicitly, we compute the fraction of $j$-type links per node, connecting two infected nodes (the cascade bulk), 
\begin{equation}
E_j^{II} (t) = \dfrac{\sum_{k,\textbf{k},\textbf{m}} P(k)P(\textbf{k}) m_j \ikm (t) }{\sum_{k,\textbf{k},\textbf{m}} P(k)P(\textbf{k}) m \ikm (t)},
\end{equation} 
and susceptible and infected nodes (cascade surface),
\begin{equation}
E_j^{SI} (t) = \dfrac{\sum_{k,\textbf{k},\textbf{m}} P(k)P(\textbf{k}) m_j \skm (t) }{\sum_{k,\textbf{k},\textbf{m}} P(k)P(\textbf{k}) m \skm (t)},
\end{equation} 
such that $\sum_j E^{II} = \sum_j E^{SI} = 1$. Now if $\Pw$ is bimodal ($n = 2$), then $E_1^{SI} + E_2^{SI} = 1$, $E_1^{II} + E_2^{II} = 1$,  and we may remove the index $j$ (Fig.~\ref{fig:2}d-e and Fig.~\ref{fig:3}c). The quantities $E_j^{II}$ and $E_j^{SI}$ diverge from $1/2$ with amplitude dependent on the absolute difference of the speed from the dynamics on unweighted networks ($\sigma = 0$) where $E_j^{II} = E_j^{SI} = 1/2$.

\subsection*{Data description}

We perform data-driven simulations of our threshold model in two large-scale, empirical social networks. The first is a network of $N=138,592$ English Wikipedia editors contributing to articles about politics. Each of the $740,397$ directed links (defining an edit, revert, restore, or vote action in an article) has a sign ($\pm$), interpreted as the parity of trust between connected editors (for free access online and details see~\cite{WikiSigned}). In our study we remove self-loops and assume bidirectional links appear as undirected links with their original sign (if they shared the same sign), while choosing a sign randomly in the case where they appear with different signs (such edges only form $0.96\%$ of the network, so their effect is not significant). Unidirectional links are also regarded as undirected with their original sign. Finally, we associate $+$ and $-$ tie signs to strong ($w_1$) and weak ($w_2$) links. The network has a broad degree distribution, a fraction $\delta = 0.88$ of strong links and average weight $\mu = 2.7$.

The second data set is an aggregated, static social network of $N=6,243,322$ individuals connected by $16,783,865$ undirected links with weights defined as the number of phone calls between people in an observation period of 6 months (a link exists if people have mutually called each other at least once). All individuals are customers of a single phone provider with $20\%$ market share in an undisclosed European country. Degree and weight distributions are broad and can be approximated by power-law and lognormal distributions, respectively (for details see~\cite{onnela2007analysis}). Since for the MPC network $\Pw$ is fixed, we introduce a method to scale $\sigma$ without changing the shape of the distribution, described as follows. We first assume that the MPC network has a weight set $W = \lbrace w_{1},\hdots,w_{|E|} \rbrace$, where $w_{i}$ is the weight of the $i$-th edge, and $|E|$ is the number of edges in the network. This set has mean and variance
\begin{equation}
\label{eq:MPCmoments}
\mu = \dfrac{1}{|E|}\sum_{i=1}^{|E|}w_{i} \hspace{.2in} \mbox{and} \hspace{.2in}
\sigma^{2} = \dfrac{1}{|E|}\sum_{i=1}^{|E|}(w_{i}-\mu)^{2}.
\end{equation}
Now we consider a new weight set $W^{\prime} = \lbrace \mu + \alpha(w_{1}-\mu),\hdots,\mu+\alpha(w_{|E|}-\mu) \rbrace$, where we have applied the transformation $w_{i}^{\prime} = \mu + \alpha(w_{i}-\mu)$, $i = 1, \ldots, |E|$, and $0 \leq \alpha \leq 1$ is a tuning parameter. The limits of this transformation give a Dirac delta distribution ($\alpha = 0$) or $\Pw$ ($\alpha = 1$). Substituting $w_{i}^{\prime}$ into the expression for $\sigma$, we see that the mean and standard deviation of the transformed weight set are $\mu^{\prime} = \mu$ and $\sigma^{\prime} = \alpha \sigma$. Then, we may obtain a new weight distribution retaining the shape of $\Pw$ by applying the transformation $w_{i}\mapsto w_{i}^{\prime}$. If $\sigma^{\prime}$ is the desired standard deviation, the required tuning parameter is $\alpha = \sigma^{\prime} / \sigma$.

\section*{Acknowledgements}
The authors gratefully acknowledge support from the SoSweet ANR project (ANR-15-CE38-0011-03), and G.I. a Visiting Fellowship from the Aalto Science Institute. They are very thankful for stimulating discussions with C. Droin, J. Kert\'esz, to A.-L. Barab\'asi for the mobile phone call dataset and M. Szell for the Pardus dataset.

\section*{Author contributions}
S.U., G.I., and M.K. designed the research and participated in writing the manuscript. S.U. derived the edge-heterogeneous approximate master equations, and G.I. derived the reduced dimension system. S.U. performed numerical experiments.




\bibliographystyle{naturemag}

\pagebreak

\begin{center}
{\LARGE \textbf{Supplementary Information}}\\[0.3cm]
\end{center}

\section{Approximate master equations on weighted networks}

In this section we justify and outline the derivation of approximate master equations (AMEs) for stochastic binary-state dynamics on weighted networks. We begin with a general derivation, and later show how this framework simplifies for monotone dynamics. We identify edge types by the value of their weights, however the formalism remains unchanged by distinguishing edge types with other link properties, such as direction or colour. The following derivation builds upon a formalism developed by Gleeson~\cite{porter2016dynamical,gleeson2013binary, gleeson2008cascades,gleeson2011high}.

\subsection{Binary-state dynamics}

The theoretical framework of the AMEs applies to stochastic binary-state dynamics on random networks, where nodes have degree $k$ with distribution $P(k)$. The network is assumed to be infinite and maximally random, meaning there is no correlation between $k$ and any other graph property. We attribute to each node in the network one of two possible states, susceptible ($S$) or infected ($I$). We denote by $m$ the number of infected neighbours of a node, with $0 \leq m \leq k$. We refer to $m$ interchangeably as the infected neighbour count or partial degree. A susceptible node of degree $k$ that has $m$ infected neighbours belongs to the set $S_{k,m}$, and an infected node to the set $I_{k,m}$. As such the network can be partitioned into a finite number of sets, assuming $P(k)$ is delimited by a minimum and maximum degree $k_{\text{min}} \leq k \leq k_{\text{max}}$.

We may further partition $S_{k,m}$ and $I_{k,m}$ by assuming a finite number of edge types within the network, distinguished by their weight. If edge weights take one of $n$ distinct values, it is instructive to introduce a weight vector $\textbf{w}=(w_{1},\hdots,w_{n})^{T}$ to store the $n$ values $w_{j}$. This could be generalised to any $n$-dimensional edge property vector. Further, we define the degree vector $\textbf{k} = (k_{1},\hdots,k_{n})^{T}$ and partial degree vector $\textbf{m} = (m_{1},\hdots,m_{n})^{T}$. Here, $k_{j}$ and $m_{j}$ respectively denote the number of neighbours and the number of infected neighbours of a node that are connected by an edge of weight $w_{j}$. These quantities are connected to the degree and partial degree via $k = \sum_{j}k_{j}$ and $m = \sum_{j}m_{j}$. Moreover, $S_{\textbf{k},\textbf{m}}$ and $I_{\textbf{k},\textbf{m}}$ denote the set of susceptible and infected nodes, respectively, that have degree vector $\textbf{k}$ and partial degree vector $\textbf{m}$. Every node in $S_{\textbf{k},\textbf{m}}$ belongs to a corresponding set $S_{k,m}$, as is the case for $I_{\textbf{k},\textbf{m}}$ and $I_{k,m}$. Finally, the size of these sets is quantified through $s_{\textbf{k},\textbf{m}}(t)$ and $i_{\textbf{k},\textbf{m}}(t)$, the fraction of nodes with degree vector $\textbf{k}$ who are susceptible or infected at time $t$, and have partial degree vector $\textbf{m}$. These quantities enumerate all possible node configurations over the course of any binary-state process. In other words, the sets $S_{\textbf{k},\textbf{m}}$ and $I_{\textbf{k},\textbf{m}}$ cannot be further partitioned, making $s_{\textbf{k},\textbf{m}}(t)$ and $i_{\textbf{k},\textbf{m}}(t)$ ideal functions of a rate equation formalism.

\begin{figure}
	\centering
	\includegraphics[scale=1]{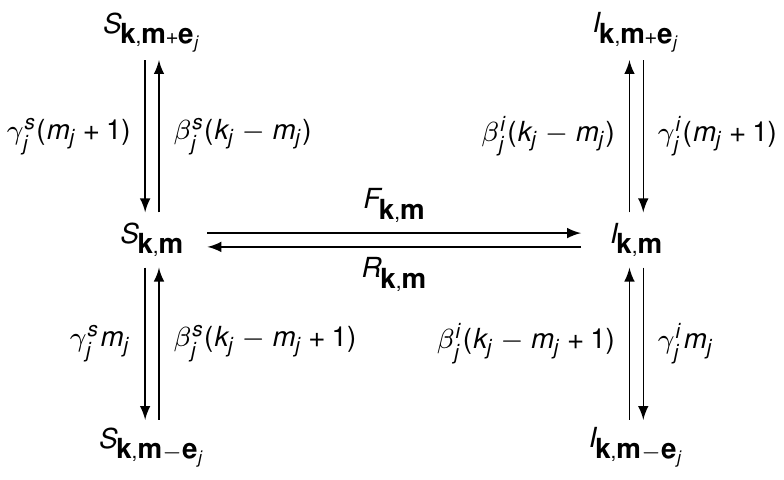}
	\caption{\small  {\bf Representation of the AME system in \esref{eqn:ame_1}{eqn:ame_2}.} The system constitutes an initial value problem, solved over a set of rate equations in $s_{\textbf{k},\textbf{m}}$ and $i_{\textbf{k},\textbf{m}}$, whose gain and loss terms, with their associated rates, are illustrated in the figure. The index $1\leq j\leq n$ enumerates the gain or loss type, corresponding to the $4n+2$ ways in which a node may enter and leave class $S_{\textbf{k},\textbf{m}}$ or $I_{\textbf{k},\textbf{m}}$. This is possible through the infection and recovery of a node's neighbours (vertical movements) and through infection and recovery of a node itself (horizontal movements).\label{fig:schem_general}}
\end{figure}

A dynamical process may be induced on such a network by assigning an initial state to each node, and allowing this state to evolve according to rates $F_{\textbf{k},\textbf{m}}$ and $R_{\textbf{k},\textbf{m}}$ per infinitesimal time step $dt$. The former is the rate of infection of susceptible nodes in $S_{\textbf{k},\textbf{m}}$ over a time interval $dt$, the latter the rate of recovery from the infected state for nodes in $I_{\textbf{k},\textbf{m}}$ over $dt$. This means all nodes sharing a degree vector $\textbf{k}$ and partial degree vector $\textbf{m}$ are equivalent in their rates of infection and recovery. The rate equations governing the sets $S_{\textbf{k},\textbf{m}}$ and $I_{\textbf{k},\textbf{m}}$ in a dynamics allowing both infection and recovery are
\begin{equation}\label{eqn:ame_1}
\begin{split}
\dfrac{d}{dt}s_{\mathbf{k},\mathbf{m}} =   -F_{\mathbf{k},\mathbf{m}} s_{\mathbf{k},\mathbf{m}}  
-\sum_{j=1}^{n} \beta_{j}^{s}(k_{j}-m_{j})s_{\mathbf{k},\mathbf{m}} + & \sum_{j=1}^{n}\beta_{j}^{s}(k_{j}-m_{j}+1)s_{\mathbf{k},\mathbf{m}-\mathbf{e}_{j}}  \\
& + R_{\mathbf{k},\mathbf{m}}i_{\mathbf{k},\mathbf{m}} -\sum_{j=1}^{n}\gamma_{j}^{s}m_{j}s_{\mathbf{k},\mathbf{m}} + \sum_{j=1}^{n}\gamma_{j}^{s}(m_{j}+1)s_{\mathbf{k},\mathbf{m}+\textbf{e}_{j}}
\end{split}
\end{equation}
and
\begin{equation}\label{eqn:ame_2}
\begin{split}
\dfrac{d}{dt}i_{\mathbf{k},\mathbf{m}} = F_{\mathbf{k},\mathbf{m}} s_{\mathbf{k},\mathbf{m}} -\sum_{j=1}^{n}\beta_{j}^{i}(k_{j}-m_{j})i_{\mathbf{k},\mathbf{m}}+ & \sum_{j=1}^{n}\beta_{j}^{i}(k_{j}-m_{j}+1)i_{\mathbf{k},\mathbf{m}-\mathbf{e}_{j}}\\
& -R_{\mathbf{k},\mathbf{m}}i_{\mathbf{k},\mathbf{m}}  -\sum_{j=1}^{n}\gamma_{j}^{i}m_{j}i_{\mathbf{k},\mathbf{m}} + \sum_{j=1}^{n}\gamma_{j}^{i}(m_{j}+1)i_{\mathbf{k},\mathbf{m}+\textbf{e}_{j}},
\end{split}
\end{equation}
where $\textbf{e}_{j}$ is the $j$-th basis vector of dimension $n$, and $\beta_{j}^{s}$, $\beta_{j}^{i}$, $\gamma_{j}^{s}$ and $\gamma_{j}^{i}$ the probabilities of a $j$-type neighbour becoming infected or recovering over an interval $dt$, calculated using the full system of $s_{\textbf{k},\textbf{m}}$ and $i_{\textbf{k},\textbf{m}}$ values. Explicitly, the $\beta_{j}$ terms quantify the rate of infection of $j$-type neighbours for both susceptible and infected nodes,
\begin{subequations}
\label{eqn:betaTerms}
\begin{align}
\beta_{j}^{s}(t) &= \dfrac{\sum_{k,\textbf{k},\textbf{m}} P(k) P(\textbf{k})(k_{j}-m_{j})F_{\textbf{k},\textbf{m}} s_{\textbf{k},\textbf{m}}(t)}{\sum_{k,\textbf{k},\textbf{m}} P(k) P(\textbf{k})(k_{j}-m_{j}) s_{\textbf{k},\textbf{m}}(t)},\\
\beta_{j}^{i}(t) &= \dfrac{\sum_{k,\textbf{k},\textbf{m}} P(k) P(\textbf{k})(k_{j}-m_{j})F_{\textbf{k},\textbf{m}} i_{\textbf{k},\textbf{m}}(t)}{\sum_{k,\textbf{k},\textbf{m}} P(k) P(\textbf{k})(k_{j}-m_{j}) i_{\textbf{k},\textbf{m}}(t)},
\end{align}
\end{subequations}
while the $\gamma_{j}$ terms give the rate of recovery of $j$-type neighbours for both susceptible and infected nodes,
\begin{subequations}
\label{eqn:gammaTerms}
\begin{align}\gamma_{j}^{s}(t) &= \dfrac{\sum_{k,\textbf{k},\textbf{m}} P(k) P(\textbf{k})m_{j}R_{\textbf{k},\textbf{m}} s_{\textbf{k},\textbf{m}}(t)}{\sum_{k,\textbf{k},\textbf{m}} P(k) P(\textbf{k})m_{j} s_{\textbf{k},\textbf{m}}(t)},\\
\gamma_{j}^{i}(t) &= \dfrac{\sum_{k,\textbf{k},\textbf{m}} P(k) P(\textbf{k})m_{j}R_{\textbf{k},\textbf{m}} i_{\textbf{k},\textbf{m}}(t)}{\sum_{k,\textbf{k},\textbf{m}} P(k) P(\textbf{k})m_{j} i_{\textbf{k},\textbf{m}}(t)},
\end{align}
\end{subequations}
where we sum over $k_{min} \leq k \leq k_{max}$, all $\textbf{k}$ such that $\sum_{j}k_{j} = k$, and all $\textbf{m}$ such that $0 \leq m_{j} \leq k_{j}$. The values $s_{\textbf{k},\textbf{m}}(t)$ and $i_{\textbf{k},\textbf{m}}(t)$ combined with the degree and degree vector distributions $P(k)$ and $P(\textbf{k})$ give us the density of infected nodes $\rho (t)$,
\begin{equation}
\rho (t) = 1 - \sum_{k,\textbf{k}, \textbf{m}} P(k) P(\textbf{k}) s_{\textbf{k},\textbf{m}}(t) = \sum_{k,\textbf{k}, \textbf{m}}P(k) P(\textbf{k}) i_{\textbf{k},\textbf{m}}(t).
\end{equation}
The initial conditions are prescribed by $i_{\textbf{k},\textbf{m}}(0)$ and $s_{\textbf{k},\textbf{m}}(0)$, subject to the normalisation condition
\begin{equation}
\sum_{\textbf{m}}i_{\textbf{k},\textbf{m}}(t) + \sum_{\textbf{m}}s_{\textbf{k},\textbf{m}}(t) = 1.
\end{equation}
As such, we have defined a closed system of deterministic equations that can be solved numerically using standard methods (\fref{fig:schem_general}).

\subsection{Monotone dynamics}

\begin{figure}[b]
	\centering
	\includegraphics[scale=1]{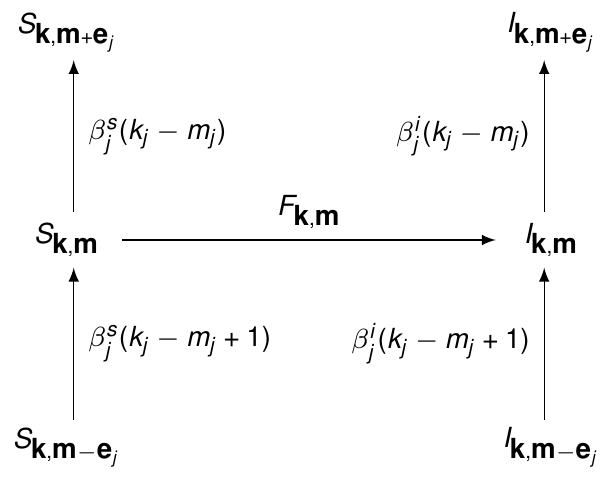}
	\caption{\small {\bf Representation of the AMEs for monotone dynamics.} A recovery rate $R_{\mathbf{k},\mathbf{m}} = 0$ implies $\gamma^{i}_{j} = \gamma^{s}_{j} = 0$, so the rate equations for $S_{\textbf{k},\textbf{m}}$ and $I_{\textbf{k},\textbf{m}}$ are characterised by only $2n + 1$ gain and loss terms, in contrast to \fref{fig:schem_general}.\label{fig:schem_mono}}
\end{figure}

The above derivation assumes generic infection and recovery rates $F_{\textbf{k},\textbf{m}}$ and $R_{\textbf{k},\textbf{m}}$. In this section, we illustrate a solution of the AMEs particular to monotone dynamics with the example of a threshold rule for complex contagion. This may be generalised to other monotone or non-recovery dynamics, where $R_{\textbf{k},\textbf{m}} = 0$. In dynamical processes on weighted networks, we are typically interested in the node properties relating to the edge weight. We define the strength of a node as the sum of edge weights across all neighbours, $q_{\textbf{k}} = \textbf{k} \cdot \textbf{w}$. Similarly, we define the partial strength as the sum of edge weights across all infected neighbours, $q_{\textbf{m}} = \textbf{m} \cdot \textbf{w}$, with $0 \leq q_{\textbf{m}} \leq q_{\textbf{k}}$. The infection rate for complex contagion can thus be expressed as 
\begin{equation}
\label{eq:thresRule}
\Fkm =
\begin{cases}
p & \quad \qm < \phi \qk \\
1 & \quad \qm \geq \phi \qk
\end{cases}, \quad k > 0,
\end{equation}
with rate of recovery $R_{\textbf{k},\textbf{m}} = 0$. Here, $p$ is the rate of spontaneous infection, whereby a susceptible node may become infected independently of the state of its neighbours. The threshold $\phi$ is the fraction of a node's total strength that must be met by the partial strength for that node to undergo induced infection. In other words, it is the fraction of a node's total received influence that must come from infected neighbours before that node itself becomes infected. Correspondingly, the master equations become (\fref{fig:schem_mono})
\begin{subequations}
\label{eqn:ame_3}
\begin{align}
\dfrac{d}{dt}s_{\mathbf{k},\mathbf{m}} &= -F_{\mathbf{k},\mathbf{m}} s_{\mathbf{k},\mathbf{m}} -\sum_{j=1}^{n}\left( \beta_{j}^{s}(k_{j}-m_{j})s_{\mathbf{k},\mathbf{m}} - \beta_{j}^{s}(k_{j}-m_{j}+1)s_{\mathbf{k},\mathbf{m}-\mathbf{e}_{j}} \right)\\
\dfrac{d}{dt}i_{\mathbf{k},\mathbf{m}} &= +F_{\mathbf{k},\mathbf{m}} s_{\mathbf{k},\mathbf{m}} -\sum_{j=1}^{n} \left( \beta_{j}^{i}(k_{j}-m_{j})i_{\mathbf{k},\mathbf{m}} - \beta_{j}^{i}(k_{j}-m_{j}+1)i_{\mathbf{k},\mathbf{m}-\mathbf{e}_{j}}\right).
\end{align}
\end{subequations}
The AME system~(\ref{eqn:ame_3}) is decoupled, so we may only consider the equation for $\skm$ when, for example, reducing the AMEs to a lower-dimensional system.

\subsection{Monotone dynamics for bimodal weight distribution}

Here we analyse the simple case of a network with arbitrary degree distribution $\Pk$ and $n = 2$ edge weights, $w_1, w_2 > 0$, which may or may not appear with equal probability in the network. The probability distribution $\Pw$ of a randomly chosen weight $w$ is
\begin{equation}
\label{eq:SI_2weightsDist}
\Pw =
\begin{cases}
\gam & \quad w = w_1 \\
1 - \gam & \quad w = w_2
\end{cases},
\end{equation}
and 0 elsewhere, with $\gam \in (0, 1)$. Assuming $w_1 \geq w_2$ for the sake of simplicity, $\gam$ is the fraction of strong edges in the network, and thus contributes to skewness in the weight distribution. The weight average and standard deviation are given by
\begin{equation}
\label{eq:SI_2weightsMS}
\mu = \gam w_1 + (1 - \gam) w_2 \quad \text{and} \quad \sigma = \sqrt{ \sum_{w}\nolimits (w - \mu)^2 P(w) } = \sqrt{ \gam (1 - \gam) } (w_1 - w_2).
\end{equation}
We may invert the linear system in \eref{eq:SI_2weightsMS} to obtain the strong and weak weights, $w_1$ and $w_2$, in terms of $\mu$ and $\sigma$,
\begin{equation}
\label{eq:SI_2typesMS}
\begin{cases}
w_1 &= \mu + \sqrt{ \frac{1 - \gam}{\gam} } \sigma \\
w_2 &= \mu - \sqrt{ \frac{\gam}{1 - \gam} } \sigma
\end{cases},
\end{equation}
where $\sigma \geq 0$ and $\mu > \sigma \sqrt{ \gam / (1 - \gam) }$~\footnote{This second condition is necessary to have positive weights only, but is not required by the following results.}. Thus, we can take $\mu$ and $\sigma$ as parameters, and use \eref{eq:SI_2typesMS} to obtain values for the weights in the network.

\begin{figure}[ht!]
	\centering
	\includegraphics[scale=1]{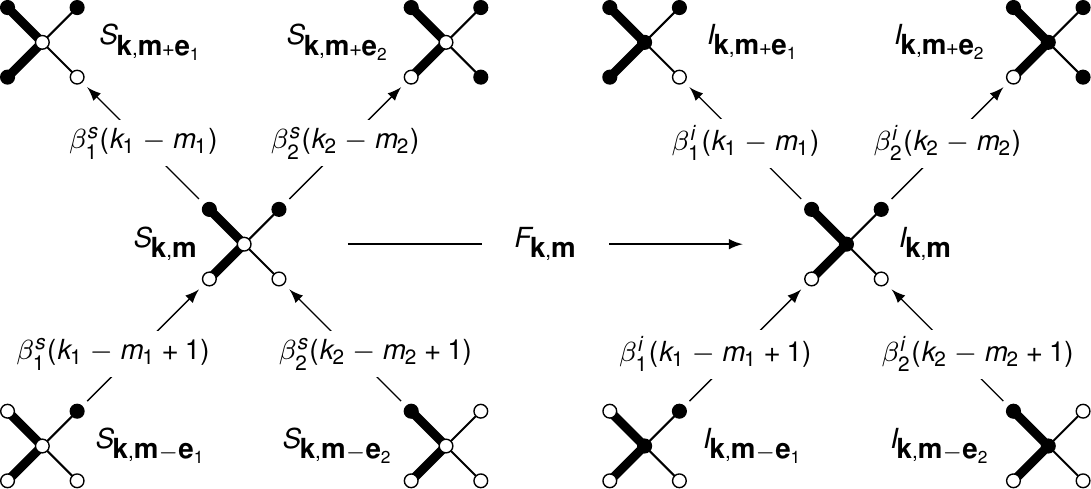}
	\vspace*{3mm}
	\caption{\small {\bf Representation of gain and loss sets for node class $\textbf{k} = (2,2)$, $\textbf{m} = (1,1)$.} Nodes in sets $S_{\textbf{k},\textbf{m}}$ and $I_{\textbf{k},\textbf{m}}$ have two possible ways of entering and exiting the class through neighbour infection. This figure corresponds to the set of $\textbf{k}=(2,2)$ nodes in \fref{fig:config_rels}. \label{fig:schem_symmetric}}
\end{figure}

As for the AME formalism in the case of a bimodal weight distribution, the weight, degree and partial degree vectors are $\wv = (w_1, w_2)^{\mathrm{T}}$, $\kv = (k_1, k_2)^{\mathrm{T}}$, and $\mv = (m_1, m_2)^{\mathrm{T}}$, respectively, subject to the constraints $k = k_1 + k_2$ and $m = m_1 + m_2$. Moreover, $\qk = k_1 w_1 + k w_2$ and $\qm = m_1 w_1 + m_2 w_2$. Due to the degree constraint and \eref{eq:SI_2weightsDist}, the degree vector takes the values $\kv = (0, k), (1, k - 1), \ldots, (0, k)$, which are binomially distributed in the network according to
\begin{equation}
\label{eq:SI_2weightsPkv}
\Pkv = \binom{k}{k_1} \gam^{k_1} (1 - \gam)^{k - k_1} = \Bkk1(\gam),
\end{equation}
Further, the sum over degrees and degree and partial degree vectors can be written explicitly as
\begin{equation}
\label{eq:SI_2weightsSums}
\sumkkm \bullet = \sum_{k=k_{min}}^{k_{max}} \sum_{k_1 = 0}^k \sum_{m_1 = 0}^{k_1} \sum_{m_2 = 0}^{k_2} \bullet.
\end{equation}
With \esref{eq:SI_2weightsPkv}{eq:SI_2weightsSums} and a given degree distribution $\Pk$, we may write explicitly the full and reduced AME systems, solve them numerically, and explore the behaviour of the fraction of infected nodes $\rho(t)$ as a function of all parameters.

\begin{figure}[ht!]
	\centering
	\includegraphics[scale=.8]{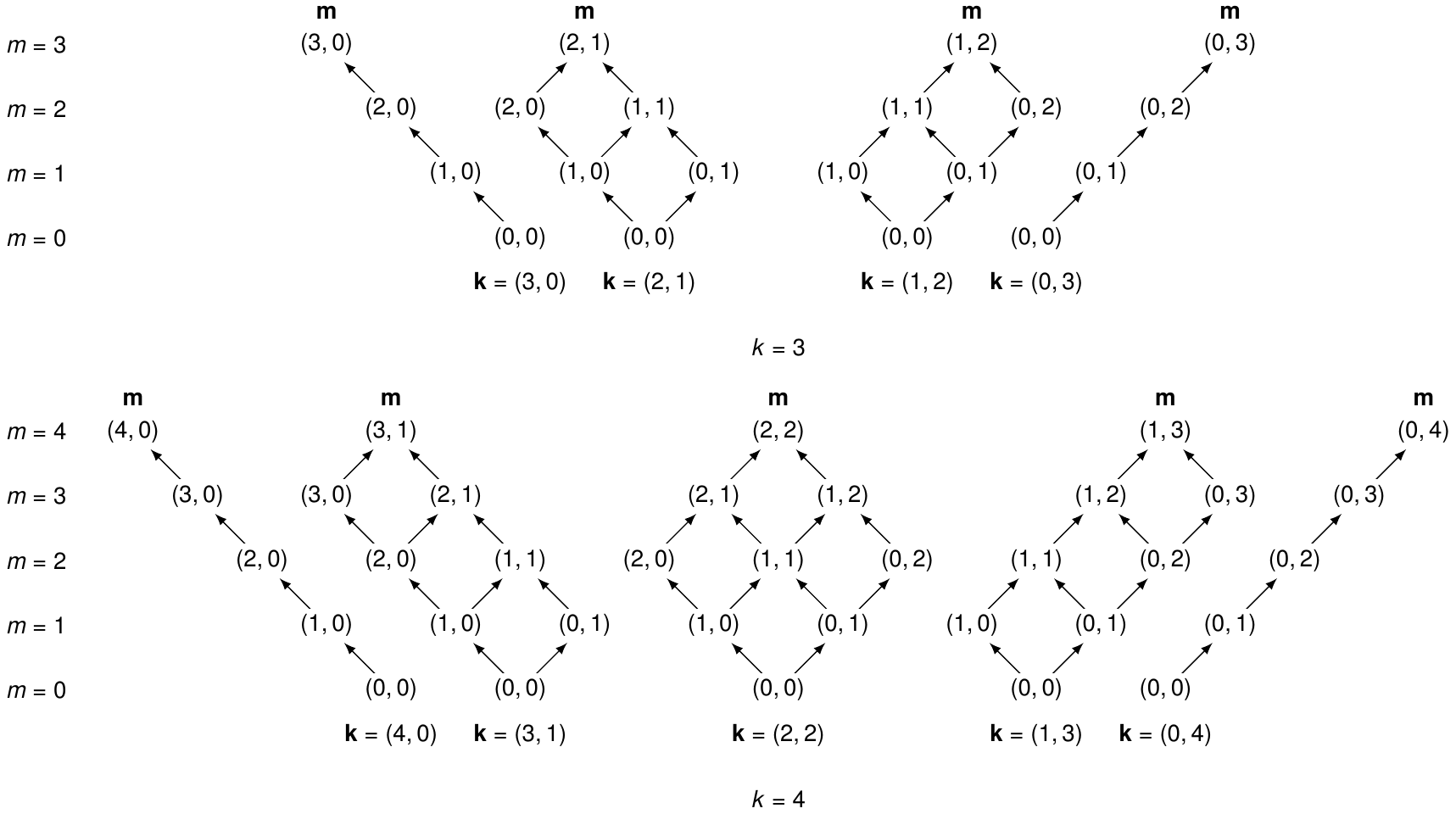}
	\caption{\small {\bf Possible $(\textbf{k},\textbf{m})$ classes for $k=3,4$ with $n=2$, and flows between them.} Note that it is impossible for a node to move between classes of different $\textbf{k}$, since the degree vector of a node is fixed in time. In the non-recovery model of the figure, it is also impossible to make a downward transition to a class with lower $m$. \label{fig:config_rels}}
\end{figure}

The bimodal case is ideal as a means of illustrating how a given node may occupy a series of $(\textbf{k},\textbf{m})$ classes over the course of a dynamical process. By taking the example of a node in class $\textbf{k} = (2,2)$, $\textbf{m} = (1,1)$ adhering to the infection rate $F_{\textbf{k},\textbf{m}}$ (\fref{fig:schem_symmetric}), we illustrate the interdependencies of various node classes and possible flows between them. This class corresponds to nodes with degree $k=4$, consisting of two strong and two weak neighbours, one of each being infected. It follows that two ways in which a node may leave this class is by an additional neighbour of either type becoming infected. Similarly, two ways in which a node may enter the class is by having only one infected neighbour of either edge type, and gaining an infected neighbour of the opposite type (\fref{fig:config_rels}). We note that although the degree vector $\textbf{k}$ of a node is fixed throughout the dynamical process, its partial degree vector $\textbf{m}$ is free to change according to the number and edge-types of its infected neighbours. This allows us to attribute relative sizes to each of the $(\textbf{k},\textbf{m})$ classes. A node's class is a dynamic quantity, and it is the flow of nodes through each class that we use to characterise the state of the system through $s_{\textbf{k},\textbf{m}}(t)$ and $i_{\textbf{k},\textbf{m}}(t)$ over time.

\subsection{Reduced AMEs}

To reduce the dimension of the weighted AMEs for monotone dynamics [\eref{eqn:ame_3}] in the case of a stepwise infection rate $\Fkm$ [\eref{eq:thresRule}], we need to consider system-wide quantities that are more aggregated than $\skm$. We take the probability $\rho(t)$ that a randomly chosen node is infected, i.e. the fraction of infected nodes in the network, and the probability $\nuj(t)$ that a randomly chosen neighbour (across a $j$-type edge) of a susceptible node is infected (see Methods). We start by proposing an exact solution for the AME system in terms of the ansatz
\begin{equation}
\label{eq:SI_AMEansatz}
\skm(t) = e^{-p t} \prodj \Bkmj [\nuj(t)]
\quad \text{for} \quad \qm < \phi \qk,
\end{equation}
where $\Bkmj = \binom{k_j}{m_j} \rho^{m_j} (1 - \rho)^{k_j - m_j}$ is the binomial distribution. The meaning of the ansatz in \eref{eq:SI_AMEansatz} is quite intuitive and takes into account two processes. First, a susceptible node with $k_j$ edges of type $j$, is connected to $m_j$ infected nodes with the binomially distributed probability $\Bkmj(\nuj)$. Second, for $\qm < \phi \qk$ a susceptible node does not fulfill the threshold rule and can only become infected spontaneously with probability $e^{-p t}$, since the system is progressively being filled due to spontaneous infection. Considering these processes as independent leads to the product in \eref{eq:SI_AMEansatz}.

The next step is to insert the ansatz~(\ref{eq:SI_AMEansatz}) into the AME system~(\ref{eqn:ame_3}) and derive a set of ordinary differential equations (ODEs) for the aggregated quantities $\rho$ and $\nuj$. Taking the time derivative $\dskm$ of \eref{eq:SI_AMEansatz} (i.e. the left-hand side of the AME system) we get
\begin{equation}
\label{eq:SI_ansatzINames1}
\dskm = \left( \sumj \left[ \frac{m_j}{\nuj} - \frac{k_j - m_j}{1 - \nuj} \right] \dnuj - p \right) \skm.
\end{equation}
Then, we use the infection rate of weighted contagion for $\qm < \phi \qk$, the ansatz~(\ref{eq:SI_AMEansatz}) and the binomial identity
\begin{equation}
\label{eq:SI_binomIdent}
\Bkmoj(\nuj) = \frac{1 - \nuj}{\nuj} \frac{m_j}{k_j - m_j + 1} \Bkmj(\nuj),
\end{equation}
in the right-hand side of the AME system to obtain
\begin{align}
\label{eq:SI_ansatzINames2}
-\Fkm \skm - \sumj \bj (k_j - m_j) \skm &+ \sumj \bj (k_j - m_j + 1) \skmej = \nonumber \\
&\left[ -p + \sumj \bj \left( m_j - k_j + \frac{1 - \nuj}{\nuj} m_j \right) \right] \skm.
\end{align}
Equating \esref{eq:SI_ansatzINames1}{eq:SI_ansatzINames2} as in the AME system, and separating terms for a given value of $j$ from the rest ($i \neq j$) leads to
\begin{equation}
\label{eq:SI_ansatzINames3}
\frac{ (1 - \nuj)m_j + \nuj (m_j - k_j) }{ \nuj } \left( \frac{\dnuj}{1 - \nuj} - \bj \right) = \sum_{i \neq j}^n \frac{ (1 - \nui)m_i + \nui (m_i - k_i) }{ \nui } \left( \bi - \frac{\dnui}{1 - \nui} \right).
\end{equation}
Since the left-hand side of \eref{eq:SI_ansatzINames3} depends on the function $\nuj$ and its derivative only, while the right-hand side depends on the rest of the functions $\nui$, both sides must be equal to some constant $c_j$. For the left-hand side, this means that
\begin{equation}
\label{eq:SI_ansatzINames4}
\frac{\dnuj}{1 - \nuj} - \bj = c_j \frac{\nuj}{m_j - \nuj k_j}, \quad \forall m_j, k_j.
\end{equation}
For the ODE~(\ref{eq:SI_ansatzINames4}) to hold regardless of the values of $m_j$ and $k_j$, we need $c_j = 0$. Then, the condition on $\nu_j$ such that the ansatz~(\ref{eq:SI_AMEansatz}) is a solution of the AME system is
\begin{equation}
\label{eq:SI_condNu}
\frac{\dnuj}{1 - \nuj} = \bj.
\end{equation}
This ODE has the initial condition $\nuj(0) = \rho(0) = 0$, obtained by evaluating \eref{eq:SI_AMEansatz} at $t = 0$ and comparing with the expression $\Bkmj (0)$, which corresponds to an infinitesimally small initial infection randomly distributed in the network (see Methods).

The next step is to extend a general result derived by Gleeson in~\cite{gleeson2013binary} [Eqs.~(F6)--(F10) therein] to the case of weighted networks. We start by multiplying the AME system~(\ref{eqn:ame_3}) by $\Pk \Pkv (k_j - m_j)$ and summing over $k$, $\kv$, and $\mv$,
\begin{align}
\label{eq:SI_GleesonCond1}
&\frac{d}{dt} \sumkkm \Pk \Pkv (k_j - m_j) \skm = - \sumkkm \Pk \Pkv (k_j - m_j) \Fkm \skm \nonumber \\
&- \sumkkm \Pk \Pkv \sumi \bi (k_j - m_j) \big[ (k_i - m_i) \skm - (k_i - m_i + 1) \skmei \big].
\end{align}
From the definition of $\bj$ in \eref{eqn:betaTerms}, the first term on the right hand side of \eref{eq:SI_GleesonCond1} may be written as
\begin{equation}
\label{eq:SI_GleesonCond2}
- \bj \sumkkm \Pk \Pkv (k_j - m_j) \skm.
\end{equation}
As for the second term on the right hand side, when $i = j$ the term telescopes to \eref{eq:SI_GleesonCond2}, and for $i \neq j$ it telescopes to 0. Overall, we can rearrange \eref{eq:SI_GleesonCond1} and obtain
\begin{equation}
\label{eq:SI_GleesonCond3}
\bj = -\frac{1}{2} \frac{d}{dt} \ln \sumkkm \Pk \Pkv (k_j - m_j) \skm.
\end{equation}
Since $\bj = - \frac{d}{dt} \ln (1 - \nuj)$ from \eref{eq:SI_condNu}, equating \esref{eq:SI_condNu}{eq:SI_GleesonCond3} implies that
\begin{equation}
\label{eq:SI_GleesonCond4}
d_j (1 - \nuj)^2 = \sumkkm \Pk \Pkv (k_j - m_j) \skm,
\end{equation}
with $d_j$ a constant that can be determined from initial conditions. Assuming an infinitesimally small fraction of infected nodes randomly distributed in the network (see Methods), and since $\nuj(0) = \rho(0) = 0$ and $\Bkmi(0) = \delta_{m_i,0}$ with $\delta_{ij}$ the Kronecker delta, we have
\begin{equation}
\label{eq:SI_GleesonCond5}
d_j = \sumkkm \Pk \Pkv (k_j - m_j) \prodi \Bkmi (0) = \sumkk \Pk \Pkv k_j = z_j,
\end{equation}
where $z_j$ is the average number of $j$-type edges a node has in the network, or average $j$-degree. Thus,
\begin{equation}
\label{eq:SI_GleesonEq}
\sumkkm \Pk \Pkv (k_j - m_j) \skm = z_j (1 - \nuj)^2.
\end{equation}

The next step is to use \eref{eq:SI_GleesonEq} to find a new expression for $\bj$ and thus write the ODE~(\ref{eq:SI_condNu}) explicitly in terms of $\nuj$. Noting that the left-hand side of \eref{eq:SI_GleesonEq} is the denominator in the definition of $\bj$, we get
\begin{align}
\label{eq:SI_betaExpl1}
\bj &= \frac{1}{z_j (1 - \nuj)^2} \left[ p \sumkk \Pk \Pkv \sumLess (k_j - m_j) \skm + \sumkk \Pk \Pkv \sumMore (k_j - m_j) \skm \right] \nonumber \\
&= \frac{1}{z_j (1 - \nuj)^2} \left[ z_j (1 - \nuj)^2 - (1 - p) \sumkk \Pk \Pkv \sumLess (k_j - m_j) \skm  \right] \nonumber \\
&= \frac{1}{1 - \nuj} \Bigg[ 1 - \nuj - (1 - p) e^{-p t} \sumkk \frac{k_j}{z_j} \Pk \Pkv \sumLess \Bkomj(\nuj) \prodinj \Bkmi (\nui) \Bigg],
\end{align}
where the sums $\sumLess$ and $\sumMore$ run over all partial degree vectors $\mv$ that comply with their respective inequalities, and we have also inserted the ansatz~(\ref{eq:SI_AMEansatz}) and the binomial identity $(k_j - m_j) \Bkmj(\nuj) = k_j (1 - \nuj) \Bkomj(\nuj)$ to simplify the expression of $\bj$. Moreover, we may introduce the {\it response function} of the monotone, threshold-driven dynamics of our model,
\begin{equation}
\label{eq:SI_respFunction}
\fkm =
\begin{cases}
0 & \quad \qm < \phi \qk \\
1 & \quad \qm \geq \phi \qk
\end{cases}, \quad k > 0,
\end{equation}
with $f(\0v, \0v) = 0$ [a function that activates when a $(\kv, \mv)$-class node fulfils the threshold condition and gets infected], in order to invert the restricted sum of \eref{eq:SI_betaExpl1},
\begin{align}
\label{eq:SI_symSums}
\sumLess \Bkomj(\nuj) \prodinj \Bkmi (\nui) &= \summ [1 - \fkm] \Bkomj(\nuj) \prodinj \Bkmi (\nui) \nonumber \\
&= 1 - \sumMore \Bkomj(\nuj) \prodinj \Bkmi (\nui).
\end{align}
Overall, comparing \esref{eq:SI_condNu}{eq:SI_betaExpl1}, we can write an explicit ODE for $\nuj$,
\begin{equation}
\label{eq:SI_nuODE}
\frac{d}{dt} \nuj = g_j(\nuv, t) - \nuj,
\end{equation}
with $\nuv = (\nu_1, \ldots, \nu_n)^{\mathrm{T}}$, $j = 1, \ldots, n$, and the function $g_j(\nuv, t)$ given by
\begin{equation}
\label{eq:SI_gFactor}
g_j(\nuv, t) = f_t + (1 - f_t) \sumkk \frac{k_j}{z_j} \Pk \Pkv \sumMore \Bkomj(\nuj) \prodinj \Bkmi (\nui),
\end{equation}
where we have defined $f_t = 1 - (1 - p) e^{-p t}$.

Even though \eref{eq:SI_nuODE} is closed and in this sense equivalent to the AME system~(\ref{eqn:ame_3}), we may also derive a corresponding ODE for $\rho$, since we are mainly interested in the temporal evolution of the fraction of infected nodes in the network. From the definition of $\rho$ and the AME system we have
\begin{align}
\label{eq:SI_rhoDeriv1}
\drho = - \sumkkm \Pk \Pkv \dskm &= \sumkkm \Pk \Pkv \Fkm \skm \nonumber \\
&\quad + \sumkkm \Pk \Pkv \sumj \bj \big[ (k_j - m_j) \skm - (k_j - m_j + 1) \skmej \big],
\end{align}
where the second term in the right-hand side telescopes to zero. Then, we use an algebraic manipulation similar to that of \eref{eq:SI_betaExpl1} to obtain
\begin{align}
\label{eq:SI_rhoDeriv2}
\drho &= p \sumkk \Pk \Pkv \sumLess \skm + \sumkk \Pk \Pkv \sumMore \skm \nonumber \\
&= 1 - \rho - (1 - p) e^{-pt} \sumkk \Pk \Pkv \sumLess \prodj \Bkmj(\nuj).
\end{align}
Thus, the ODE for $\rho$ is
\begin{equation}
\label{eq:SI_rhoODE}
\frac{d}{dt} \rho = h(\nuv, t) - \rho,
\end{equation}
where the function $h(\nuv, t)$ is given by
\begin{equation}
\label{eq:SI_hFactor}
h(\nuv, t) = f_t + (1 - f_t) \sumkk \Pk \Pkv \sumMore \prodj \Bkmj(\nuj).
\end{equation}

Combining all of these results, the AME system~(\ref{eqn:ame_3}) is reduced to a closed system of $n+1$ coupled, non-linear ODEs,
\begin{subequations}
\label{eq:SI_reducedAMEs}
\begin{align}
\dnuj &= g_j(\nuv, t) - \nuj, \\
\drho &= h(\nuv, t) - \rho,
\end{align}
\end{subequations}
with the quantities $g_j(\nuv, t)$ and $h(\nuv, t)$ given explicitly by \esref{eq:SI_gFactor}{eq:SI_hFactor}.

\section{Combinatorial solution of parameter space boundaries}

\begin{figure}[t]
	\centering
	\includegraphics[scale=0.99]{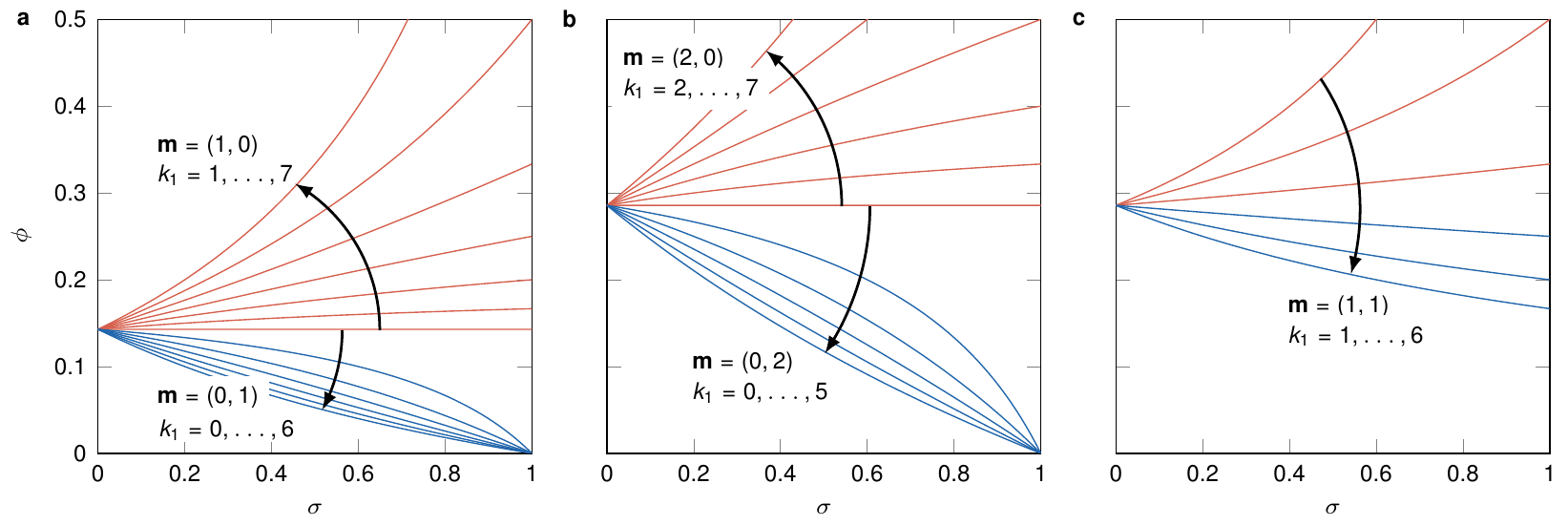}
	\caption{\small {\bf Phase boundaries in $(\sigma, \phi)$-parameter space for a $k$-regular random network ($k = 7$) with $m=1, 2$.} \textbf{(a)} Boundaries of regions where just one infected neighbour of type $j=1, 2$ is sufficient to induce infection. Curves in red indicate that the associated $(\textbf{k},\textbf{m})$ class produces a speed-up effect on the spreading process relative to the same process on an unweighted network. Conversely, classes associated with the curves in blue produce a slow-down effect on cascades. \textbf{(b)} Similar boundaries for the networks where two infected neighbours of the same type are sufficient to cause induced infection, over a range of degree vectors. \textbf{(c)} Boundaries where one infected neighbour of each type causes infection.\label{fig:combinatorial_arg}}
\end{figure}

The dynamics of threshold driven contagion on weighted networks depends on the stepwise infection rate $\Fkm$ of \eref{eq:thresRule}. Considering the case of equality, $\qm = \phi \qk$, and writing $q_{\textbf{k}}$ and $q_{\textbf{m}}$ explicitly, we obtain $\phi = \textbf{m}\cdot\textbf{w} / \textbf{k}\cdot\textbf{w}$. Noting that the $\sigma$ dependence is contained in the weight vector $\textbf{w}$, we solve for $\textbf{k}$ and $\textbf{m}$, and associate the solution with a unique boundary in $(\sigma, \phi)$-parameter space separating regions of differing $t_{r}$, the relative time of cascade emergence (\fref{fig:combinatorial_arg}). In other words, boundaries for $t_{r}$ in $(\sigma, \phi)$-parameter space separate network configurations where the corresponding $(\textbf{k},\textbf{m})$ class does and does not satisfy the threshold rule $\qm \geq \phi \qk$, thus promoting or hindering spreading. In \fref{fig:combinatorial_arg} we enumerate all possible boundaries for up to two infected neighbours in the case of a $k$-regular random network ($k = 7$) and a bimodal weight distribution ($n=2$). \fref{fig:combinatorial_arg}a shows the case where one strong infected neighbour, $\textbf{m}=(1,0)$, is sufficient to cause infection for nodes with $k_{1}=1,\hdots,k$ strong neighbours. These curves are shown in red, since the corresponding node classes induce a faster cascade of spreading compared to the same process carried out on an unweighted network. Since the weight vector is $\textbf{w}=(\mu + \sigma, \mu - \sigma)^{T}$ for weight mean $\mu$ and skewness $\delta = 0.5$, boundaries can be written explicitly as
\begin{equation}
\phi = \dfrac{m_{1}w_{1} + m_{2}w_{2}}{k_{1}w_{1} + k_{2}w_{2}} = \dfrac{\mu (m_1 + m_2) + \sigma (m_1 - m_2)}{\mu (k_1 + k_2) + \sigma (k_1 - k_2)}.
\end{equation}
Curves in blue are the boundaries where one weak infected neighbour, $\textbf{m}=(0,1)$, is sufficient to induce infection (and a slower cascade than the unweighted case). These curves are enumerated by the number of strong neighbours, $k_{1}=0,\hdots,k-1$.
Curves in \fref{fig:combinatorial_arg}b are analogous to \fref{fig:combinatorial_arg}a, except replacing $\textbf{m}=(1,0)$ and $\textbf{m}=(0,1)$ with $\textbf{m}=(2,0)$ and $\textbf{m}=(0,2)$. Finally, \fref{fig:combinatorial_arg}c corresponds to the boundaries due to node sets with $\textbf{m}=(1,1)$, where having two infected neighbours, one of each type, is sufficient for these classes to undergo induced infection.

\section{Comparison of numerical experiment and AME solutions}

\begin{figure}[t]
    \centering
    \begin{subfigure}[b]{0.45\textwidth}
        \includegraphics[width=\textwidth]{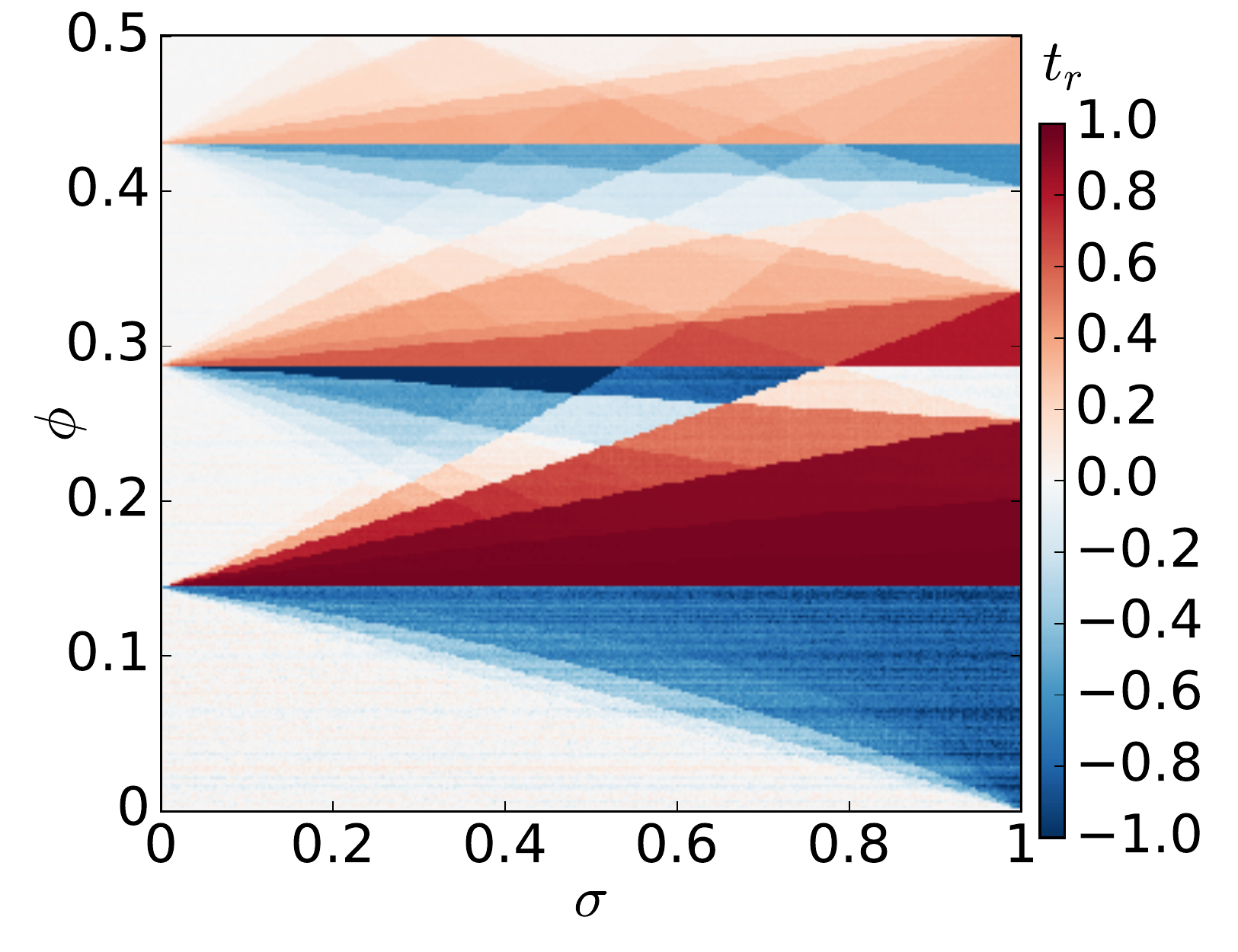}
        \caption{Numerical simulation}
    \end{subfigure}
    \begin{subfigure}[b]{0.45\textwidth}
        \includegraphics[width=\textwidth]{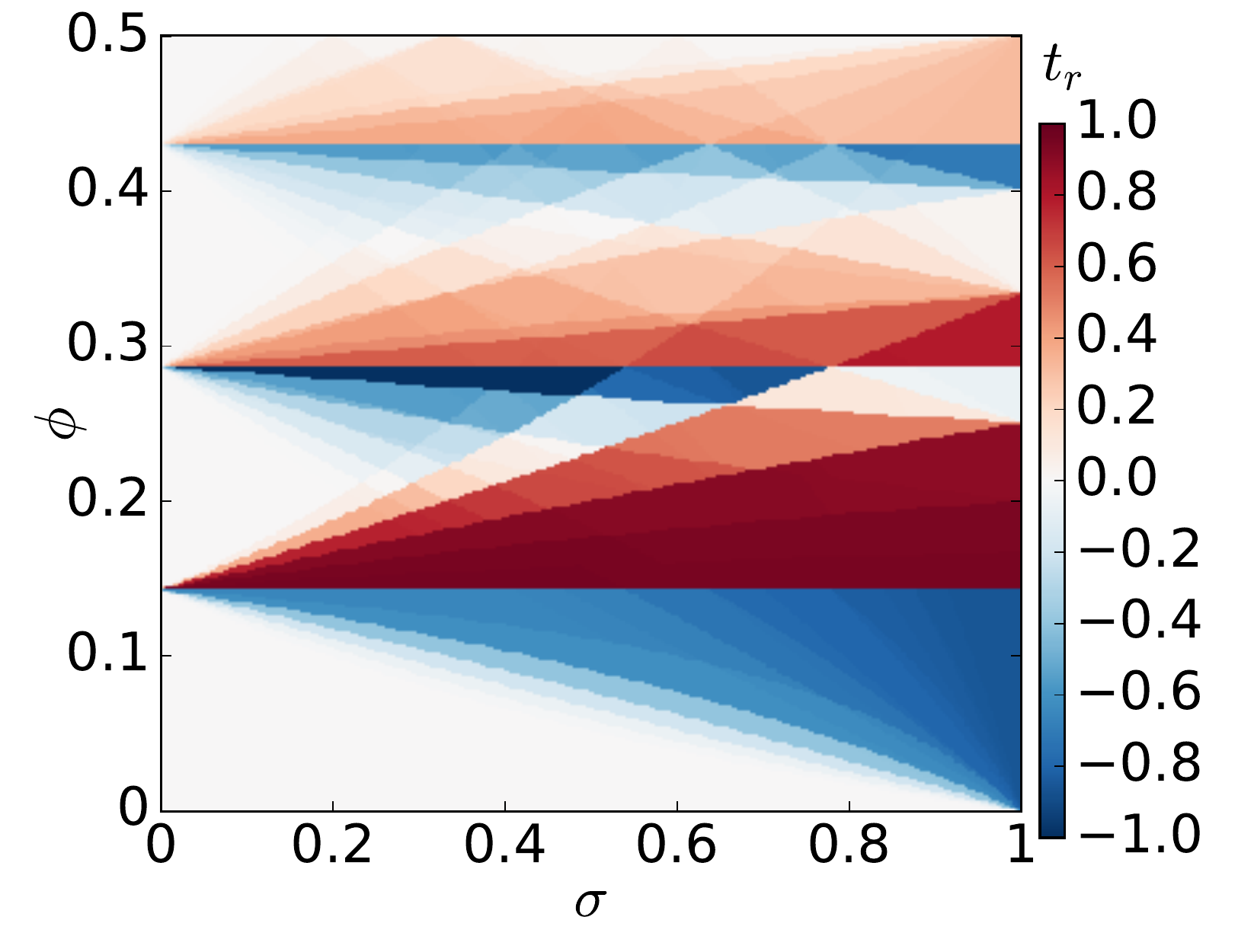}
        \caption{AME solution}
    \end{subfigure}
    \caption{\small {\bf Comparison between numerical simulations and AME solutions.}  $(\sigma, \phi)$-parameter space for the relative time $t_r$ of cascade emergence, obtained by Monte Carlo numerical simulations (a) and the numerical solutions of the full and reduced AME systems (b), the last two of which are indistinguishable. Numerical simulations consider $k$-regular random networks ($k=7$) with $N=10^4$, $p=2 \times 10^{-4}$, and averages over $25$ realisations.}
    \label{fig:Sparsp}
\end{figure}

As discussed in the main text, the behaviour of threshold driven contagion over weighted networks, in Monte Carlo simulations or as predicted by numerically computing AME solutions, is remarkably consistent. We have illustrated this similarity by comparing $t_r$ values over $(\sigma, \phi)$-parameter space, as well as by plotting the temporal evolution of the infection density $\rho$ and other quantities characterising the dynamics. To further compare the $(\sigma, \phi)$-parameter space between Monte Carlo simulations and AME solutions, we show them side-by-side in \fref{fig:Sparsp} and quantify their similarity by computing the mean absolute difference
\begin{equation}
MD[t_r(\sigma,\phi)] = \frac{\sum_{\sigma}\sum_{\phi}|t^{\text{sim}}_r(\sigma,\phi)-t^{\text{theo}}_r(\sigma,\phi)|}{N_{\sigma}N_{\phi}},
\end{equation}
where $N_{\sigma}$ and $N_{\phi}$ are the number of points considered in each dimension of the parameter space, and $t_r^{\text{sim}}(\sigma,\phi)$ and $t_r^{\text{theo}}(\sigma,\phi)$ are the relative times of cascade emergence for a given $(\sigma, \phi)$ point, measured by numerical simulations or AME solutions. This quantity is very small, $MD=2.8\times 10^{-7}$, which indicates that even if we have taken several simplifying assumptions during the derivation of the full and reduced AME systems [\esref{eqn:ame_3}{eq:SI_reducedAMEs}], they provide an extremely good approximation of the spreading process with differences only due to small statistical fluctuations in finite-size numerical simulations.

\section{Other heterogeneous synthetic and real networks}

\begin{figure}[t]
    \centering
    \begin{subfigure}[b]{0.45\textwidth}
        \includegraphics[width=\textwidth]{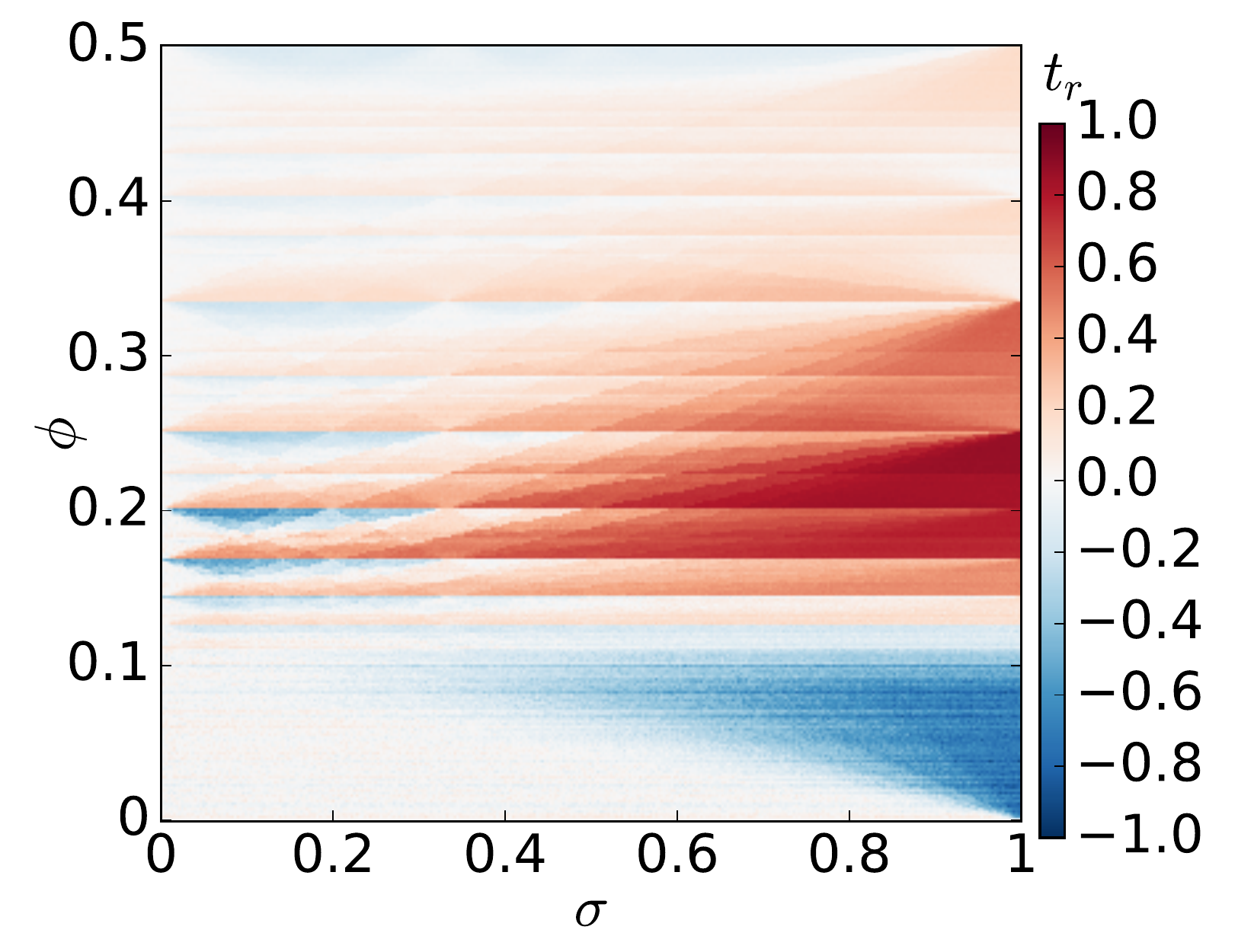}
        \caption{Random network}
    \end{subfigure}
    \begin{subfigure}[b]{0.45\textwidth}
        \includegraphics[width=\textwidth]{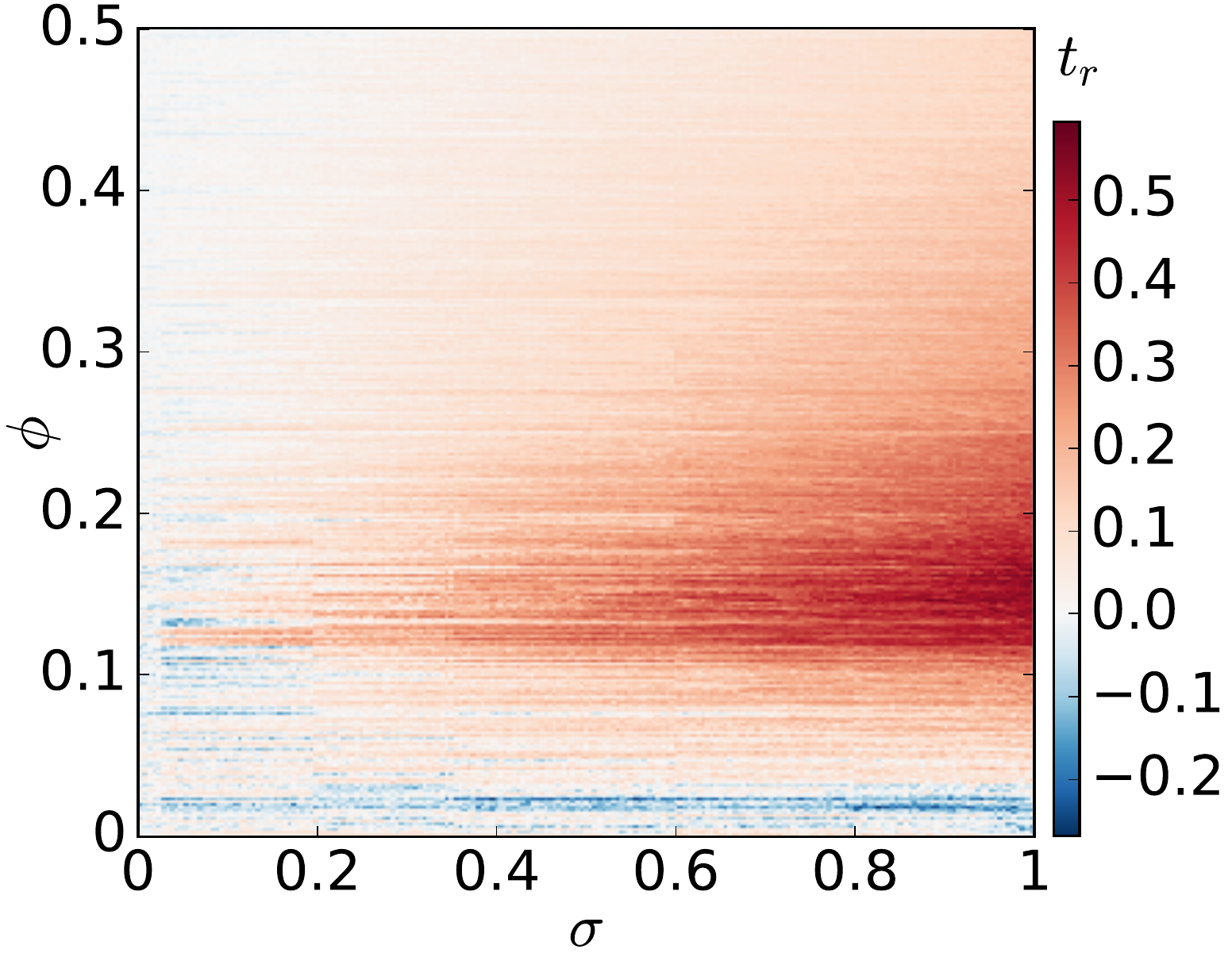}
        \caption{Pardus social network}
    \end{subfigure}
    \caption{\small {\bf Other heterogeneous synthetic and real networks.} $(\sigma, \phi)$-parameter space for the relative time $t_r$ of cascade emergence, simulated on a configuration-model random network (a) and the Pardus signed social network (b). Numerical simulations on (a) consider networks with $N=10^4$, average degree $z=7$ and averages over $25$ realisations. All simulations correspond to $p=0.0002$.}
    \label{fig:Sgen}
\end{figure}

In the main text we have seen that, even for heterogeneous synthetic and real world networks, threshold driven contagion strongly depends on link weights via simple mechanisms that can be understood by master equations or combinatorial arguments, and develops spreading cascades that are either faster of slower than their counterparts in unweighted contagion, depending on the values of $\sigma$ and $\phi$. Here we further support this argument by exploring two additional examples of synthetic and empirical networks. The synthetic structure is a configuration-model random network with average degree $z = 7$, and bimodal weight distribution with average $\mu=1$ and skewness $\delta = 0.5$ (\fref{fig:Sgen}a). The $(\sigma, \phi)$-parameter space for $t_r$ is qualitatively very similar to the ones observed for configuration-model $k$-regular or scale-free networks, with several fast and slow cascade regimes that start from values on the $\phi$ axis determined by the harmonic series of degrees present in the network. The empirical structure is a signed social network, the alliance/enemy network of the Pardus massive multiplayer online game~\cite{szell2010multirelational}. This network consists of $N=4650$ nodes connected by $66,580$ links, of which a fraction $\delta = 0.64$ have a positive alliance sign (and are considered as strong ties by us), while the rest of the links have a negative enemy sign and are interpreted as weak. The $(\sigma, \phi)$-parameter space for $t_r$ (Fig.~\ref{fig:Sgen}b) is somewhat less structured than in the other explored networks, but it still shows regions of fast and slow cascades with respect to the unweighted case.

\end{document}